\documentclass{article}       
\usepackage{graphicx}
\newcommand{\pdifft}[1]{\frac{\partial #1}{\partial t}}
\newcommand{\pdiffx}[1]{\frac{\partial #1}{\partial x}}
\usepackage{tabularx}
\newcolumntype{L}{>{\raggedright\arraybackslash}X}
\usepackage{subfig}
\usepackage{color}
\usepackage[dvipsnames]{xcolor}

\usepackage{amsmath}
\usepackage{amssymb}
\usepackage{authblk}

\providecommand{\keywords}[1]
{
  \small	
  \textbf{\textit{Keywords}} #1
}

\providecommand{\subclass}[1]
{
  \small	
  \textbf{\textit{Subject class (MSC 2020)}} #1
}

\title{Leadership through influence: what mechanisms allow leaders to steer a swarm?}
\author[1]{Sara Bernardi\footnote{Corresponding author: \texttt{sara.bernardi@polito.it}}}
\author[2]{Raluca Eftimie}
\author[3]{Kevin J. Painter}
\affil[1]{Dipartimento di Scienze Matematiche (DISMA), Politecnico di Torino, Corso Duca degli Abruzzi 24, 10129 Torino, Italy}
\affil[2]{Laboratoire de math\'{e}matiques de Besan\c{c}on, UMR-CNRS 6623
Universit\'{e} de Bourgogne Franche-Comt\'{e}, 16 Route de Gray, 25000 Besan\c{c}on, France}
\affil[3]{Dipartimento Interateneo di Scienze, Progetto e Politiche del Territorio (DIST), Politecnico di Torino, Viale Pier Andrea Mattioli, 39 10125 Torino, Italy}

\date{}                     
\begin{document}





\maketitle

\begin{abstract}
Collective migration of cells and animals often relies on a specialised set of ``leaders'', whose role is to steer a population of naive followers towards some target. We formulate a continuous model to understand the dynamics and structure of such groups, splitting a population into separate follower and leader types with distinct orientation responses. We incorporate ``leader influence'' via three principal mechanisms: a bias in the orientation of leaders according to the destination, distinct speeds of movement and distinct levels of conspicuousness. Using a combination of analysis and numerical computation on a sequence of models of increasing complexity, we assess the extent to which leaders successfully shepherd the swarm. 
While all three mechanisms can lead to a successfully steered swarm, parameter regime is crucial with non successful choices generating a variety of unsuccessful attempts, including movement away from the target, swarm splitting or swarm dispersal.
\end{abstract}

\noindent \keywords{Collective migration, Follower-leader, Swarming, Nonlocal PDEs}

\noindent \subclass{92D40, 92C15}

\section{Introduction}

\noindent
\normalsize  
Collective migration underlies numerous processes, including the migration of cells during morphogenesis and cancer progression \cite{friedl2009collective,friedl2012classifying}, social phenomena such as pedestrian flow and crowding \cite{helbing1995social,kretz2006experimental,colombi2017discrete}, and the coordinated movements of animal swarms, flocks and schools \cite{dingle2014migration,westley2018collective}.

In many cases, effective migration may demand the presence or emergence of {\em leaders}, for example as an evolved strategy for herding the population to a certain destination, finding better environments, hunting or escaping {\em etc}. At the cellular level examples include epithelial wound healing, where a set of so-called leader cells at the tissue boundary appear to guide a migrating cell group \cite{vishwakarma2018mechanical}, embryonic neural crest cell invasion, where trail-blazing pioneers lead followers in the rear \cite{schumacher2016}, and kidney morphogenesis, where the lumen forms as leading cells leave the epithelialised tube in their wake \cite{atsuta2015fgf8}. Collective invasion of breast cancer appears to be driven by a specialised population defined by their expression of basal epithelial genes \cite{cheung2013collective}.

Leadership is also found in various migrating animal groups, for example arising from a cohort of braver or more knowledgeable individuals: faced by poor feeding grounds, post-reproductive females take on an apparent leadership role in the guidance of a killer whale pod, their experience offering a reserve of ecological knowledge, \cite{brent2015ecological}. As our principal motivation we consider honeybee swarms, which form as a colony outgrows its nest site. At this point the queen and two-thirds of the colony depart (leaving a daughter to succeed her) and temporarily bivouac nearby, for example on a tree branch. Over the following hours to days, a relatively small subpopulation of \emph{scout bees} ($\sim3-5\%$ of the 10,000+ strong swarm) scour the surroundings for a suitable new nest location, potentially several kilometres distant. The quality of a potential site is broadcast to other swarm members and,  once consensus is obtained, the entire colony moves to the new dwelling. Consequently, guidance of 1000s of naive insects (including the queen) is entrusted to a relatively small number of informed scouts \cite{Seeley}. Observations suggest that scouts perform a sequence of high-velocity movements towards the nest site through the upper swarm \cite{beekman2006does,schultz2008,greggers2013scouts,Seeley}, ``streaking'' that conceivably increases their conspicuousness and communicates the nest direction.



Understanding the collective and coordinated dynamics of migrating groups demands analytical reasoning. 
The mathematical and computational literature in this field encompasses a particularly wide range of approaches. Microscopic, agent-based or individual-based models describe a group as a collection of individual agents, where the evolution of each particle is tracked over time.  Benefiting from their capacity to provide a quite detailed description of an agent's dynamics, they offer a relatively natural tool to investigate collective phenomena (see, for instance, \cite{couzin2005effective,couzin2002collective,diwold2011deciding,fetecau2012mathematical,janson2005honeybee,ioannou2015potential,bernardi2018particle}).

However, as the number of component individuals become large (as would be typical for many cancerous populations, large animal groups {\em etc}), microscopic methods become computationally expensive and macroscopic approaches may become necessary. Various continuous models have been proposed to understand the collective migration dynamics of interacting populations, with {\em nonlocal PDE} frameworks becoming increasingly popular; models falling into this class have been developed in the context of both ecological and cellular movement, e.g. see \cite{mogilner1999,armstrong2006,topaz2006,Raluca,eftimie2012}. Their nonlocal nature stems from accounting for the influence of neighbours on the movements of an individual, and their relative novelty has also become a source of significant mathematical interest (see \cite{chen2020} for a review).

The aim of this paper is to investigate the impact of informed leaders on naive followers, using a nonlocal PDE model that builds on the hyperbolic PDE approach developed in \cite{Raluca}. In particular, we will explore the extent to which the presence of leaders  can result in a {\em steered swarm}, defined as a population acquiring and maintaining a {\em spatial compact profile} that is {\em consistently steered} towards a target known only to the leaders. Motivated by real-world case studies (in particular, bee swarming as described above) we assume leaders attempt to influence the swarm using one or more of three mechanisms: (i) leaders preferentially choose the direction of the target; (ii) leaders move more quickly when moving towards the target; (iii) leaders alter their conspicuousness according to the target direction. In Section 2, we introduce the full follower-leader model, along with two simple submodels -- a leader-only and a follower-only system -- designed to reveal insights into the behaviour of the full system. Section 3 explores the dynamics of the submodels, via a combination of linear stability and numerical simulation. Section 4 subsequently addresses the full system, in particular the effectiveness of the different biases. We conclude with a discussion and an outlook of future investigations. 

\section{Follower-leader swarm model} 

We assume a heterogeneous swarm composed from distinct populations of knowledgeable leaders and naive followers. Both orient according to their interactions with other swarm members, as detailed below, but leaders have ``knowledge'' of the target and therefore the direction in which the swarm should be steered. For convenience we will restrict here to one space dimension, assume fixed speeds and account for direction through separately tracking positively ($+$) and negatively ($-$) oriented populations. Without loss of generality we assume the leaders aim to herd the swarm in the ($+$) direction, influencing via: 
\begin{itemize}
    \item {\bf Bias 1, orientation.} Leaders preferentially choose the target direction.
    \item {\bf Bias 2, speed.} Leaders alter their speed according to the target direction.
    \item {\bf Bias 3, conspicuousness.} Leaders alter their conspicuousness according to the target direction.
\end{itemize}

Setting $u^{\pm}(x,t)$ and $v^{\pm}(x,t)$ to respectively denote the densities of followers and leaders at position $x\in \Omega \subset \mathbb{R}$ and time $t\in[0,\infty)$, the governing equations are 
as follows:
\begin{eqnarray} \label{model}
\pdifft{u^+} + \gamma \pdiffx{u^+} & = & -\lambda^{u^+} u^+ +\lambda^{u^-} u^-\,, \nonumber\\
\pdifft{u^-} - \gamma \pdiffx{u^-} & = & +\lambda^{u^+} u^+ -\lambda^{u^-} u^-\,, \nonumber \\
\pdifft{v^+} + \beta_+ \pdiffx{v^+} & = & -\lambda^{v^+} v^+ +\lambda^{v^-} v^-\,, \nonumber \\
\pdifft{v^-} - \beta_- \pdiffx{v^-} & = & +\lambda^{v^+} v^+ -\lambda^{v^-} v^-\,, \nonumber \\
u^\pm(x,0) & = & u_0^\pm(x)\,, \nonumber \\
v^\pm(x,0) & = & v_0^\pm(x)\,.
\end{eqnarray}
In its general setting the model is formulated under the assumption of an infinite 1D line. For the simulations later we consider a bounded interval $\Omega = [0,L]$, but wrapped onto the ring (periodic boundary conditions) to minimise the influence of boundaries. Initial conditions will be specified later.

In the above model, followers move with a fixed speed (set at $\gamma$). Leaders have potentially distinct speeds, $\beta_{\pm}$, according to whether {\bf bias 2} is in operation; for example, in the example of bee swarming, scouts engage in streaking and increase their speed when moving towards the new nest site \cite{Seeley}. 
Switching between directions is accounted for via 
the right hand side terms, where $\lambda^{u^+}$ denotes the rate at 
which a follower ($u$) turns from ($+$) to ($-$), with similar definitions for $\lambda^{u^-}, \lambda^{v^\pm}$. Note that the current model excludes switching between follower and leader status, although it is of course possible to account for such behaviour through additional role-switching transfer functions. 

\begin{figure}[t!]
    \centering
    \includegraphics[width=\textwidth]{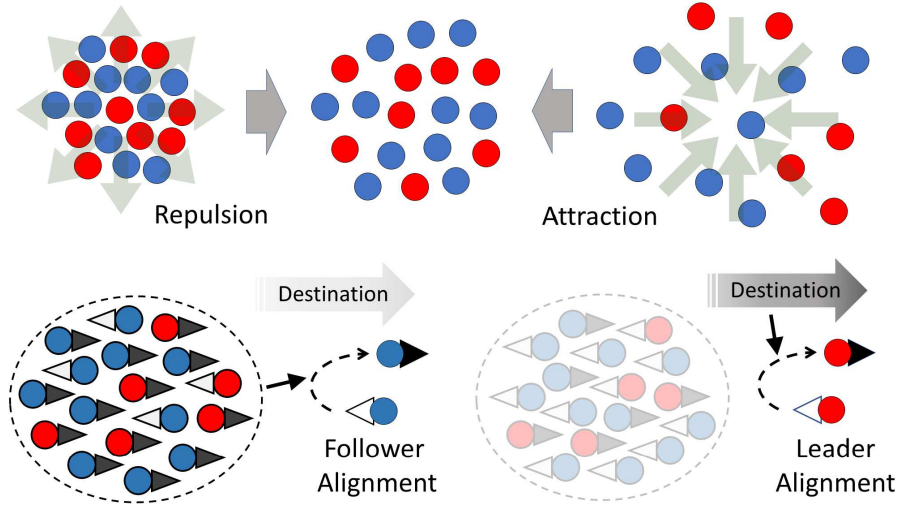}
    \caption{Assumptions underlying the turning behaviour of swarm members. Top row: attraction and repulsion are assumed to act equally on followers (blue circles) and leaders (red circles). Repulsion acts over shorter ranges, pushing individuals away from each other if they are too close; attraction acts over larger distances, pulling individuals together if they become too separated. Bottom row: alignment is distinct for followers and leaders. Followers do not know the target but are influenced by the orientation of the oncoming swarm, reorienting when they perceive the oncoming swarm is moving in the opposite direction. Leaders ignore the alignment of the swarm, biasing instead according to the target direction.}
    \label{schematic}
\end{figure}

The turning rate functions are based on interactions between swarm members where, accounting for the ``first principles of swarming'' \cite{Carrillo}, we combine {\em repulsion} (preventing collision between swarm members), {\em attraction} (preventing loss of contact and swarm dispersal) and {\em alignment} (choosing a direction according to those assumed by neighbours and/or external bias). Figure \ref{schematic} summarises the general principals upon which the model is founded.

The turning rate functions have the following form:
\[
\lambda^{i^\pm} = \lambda_1 + \lambda_2 f(y^{i^\pm})\,,
\]
for $i\in\left\{u,v\right\}$ and where $f(y) = 0.5+0.5\tanh (y-y_0)$. This assumes the turning rate smoothly and monotonically increases 
from a baseline to maximum value according to the level of {\em perceived signal}, measured separately 
for ($+$) and ($-$) follower and leader populations in $y^{u^\pm}$ and $y^{v^\pm}$. If $y_0$ is chosen in such a way that $f(0) \ll 1$ the coefficients $\lambda_1$ and $\lambda_2$ can be regarded as the {\em baseline turning rate} and the {\em highly biased turning rate} respectively. For positively-moving followers at position $x$ and time $t$, $y^{u^+}(x,t)$ combines the repulsive, attractive and alignment interactions with their neighbours into a single measure that dictates the turning rate, with similar interpretations
for $y^{u^-}$, $y^{v^{\pm}}$. Specifically, we set
\begin{equation}\label{perceived_signal}
y^{u^\pm} = Q_r^{u^{\pm}} + Q_a^{u^\pm} + Q_{l}^{u^\pm}
\end{equation}
and similarly for $y^{v^\pm}$. $Q_{r},  Q_{a}$ and $Q_{l}$ 
integrate the perceived positional and directional information from neighbours located at a distance $s \in (0, \infty)$ from the generic individual placed at $(x,t)$.

For simplicity we will assume here that followers and leaders are only distinguished by their alignment response: repulsion/attraction are taken as ``universal'' and act to keep the overall population together and avoid collisions. A noteworthy consequence of this is that a leader is not {\em bound} to choose the direction of the target: for example, if there is a danger of losing contact with the swarm the leader should be inclined to return to the fold. We adopt the following standard choices.
\begin{eqnarray*}
\label{Qrep}
Q_{r}^{u^\pm} = Q_{r}^{v^\pm} & = & q_r \int_{0}^{\infty} K_r (s) \left(u(x\pm s)+v(x\pm s)-u(x \mp s)-v(x\mp s) \right) ds\,, \\
Q_{a}^{u^\pm} = Q_{a}^{v^\pm} & = & -q_a \int_{0}^{\infty} K_a (s) \left(u(x\pm s)+v(x\pm s)-u(x \mp s)-v(x\mp s) \right) ds\,. \label{Qattr}
\end{eqnarray*}
In the above, $K_i(s)$, $i=\left\lbrace a,r \right\rbrace$, denote interaction kernels and parameters $q_a$ and $q_r$ represent the magnitude of the attraction and repulsion contributions, respectively. The attractive and repulsive terms depend on the total density of the cohort at a certain position, regardless of flight orientation, i.e. $u(x \pm s, t) = 
u^+(x \pm s, t)+u^-(x \pm s, t)$ and similarly $v(x \pm s, t) = v^+(x \pm s, t)+v^-(x \pm s, t)$. For an individual flying in the direction of a large swarm (i.e. towards overall higher total population densities), the contribution to $y$ from $Q_{r}$ will be positive (hence, an increased likelihood of turning away) and from $Q_{a}$ will be negative (hence less likely to turn away). Whether the combined contribution is then positive or negative depends on the individual parameters and the precise shape of the total density distribution.

The alignment contribution is of the general form
\begin{equation}
\label{Qal}
Q_{l}^{i^\pm} = q_{l} \int_0^\infty K_{l}(s)P^i(u^\pm,v^\pm)ds,
\end{equation}
for $i \in \left\{u,v\right\}$ and where $K_{l}(s)$ and $q_{l}$ respectively denote the alignment kernel and the magnitude of the synchronization. The functions $P^u(u^\pm,v^\pm)$ and $P^v(u^\pm,v^\pm)$ respectively represent how the swarm influences alignment for the follower and leader populations. 
Choices for $Q_{l}$, i.e. the specification of $P^i(u^\pm,v^\pm)$, form the point of distinction for the various models and are  described below, see Table \ref{table_interactions} for a summary of the models interactions. As we will see in Section \ref{leader-only_model}, the latter may simply take into account a fixed preferred direction, i.e. modeling a case where a population knows where it wants to go. 

Interaction kernels are given by the following translated Gaussian functions
\begin{equation}
K_{i}(s)= \frac{1}{\sqrt{2 \pi m_{i}^2}} \exp \left(\frac{-(s-s_{i})^2}{2m_{i}^2} \right), \quad i=r,a, l  \quad s \in [0,\infty),
\end{equation}
where $s_r$, $s_a$ and $s_{l}$ are half the length of the repulsion, attraction and alignment ranges, respectively. The constants $m_{i}$, $i=r,a, l$, are chosen to ensure $>98\%$ of the support of the kernel mass falls inside $[0,\infty)$ (specifically, $m_{i}=\frac{s_{i}}{8}$, $i=r,a, l$). This allows a high level approximation of the integral defined on $[0, \infty)$ to that defined on the whole real line.

\subsection{Follower-leader model}
\label{full_model}
The full follower-leader model assumes the following leader alignment
\begin{equation}\label{Qlv_full_model}
Q_{l}^{v^\pm} = \mp 2 q_{l} \int_0^{\infty} K_{l} (s) \varepsilon\, ds = \mbox{constant}
\end{equation}
where we call $\varepsilon$ the {\em orientation bias parameter}. Leaders ignore other swarm members for alignment, receiving instead a (spatially uniform and constant) alignment bias if the orientation bias is operating. Invoking the honeybees example, scouts have generally agreed on the new nest at swarm take-off. Generalisations could include letting $\varepsilon$ explicitly depend on a variable factor or including an influence of alignment from other swarm members.

Alignment of followers is taken to be
\begin{equation}\label{Qlu_full_model}
Q_{l}^{u^\pm} = q_{l} \int_0^{\infty} K_{l} (s) \left( u^{\mp} (x \pm  s) + \alpha_{\mp} v^{\mp} (x \pm  s) - u^{\pm} (x\mp s) - \alpha_{\pm} v^{\pm} (x\mp s) \right) ds\,.
\end{equation}
This dictates that a follower will be more likely to turn when it detects, within the region into which is moving, a large number of individuals moving in the opposite direction. Other plausible choices can be considered, however we choose the present form for its consistency with that assumed in \cite{Raluca}. Note that $\alpha_{\pm}$ are weighting parameters that distinctly weight the leader conspicuousness, {\bf bias 3}. Completely inconspicuous leaders would correspond to $\alpha_{\pm} = 0$ while if leaders are completely indistinguishable from followers $\alpha_{\pm} = 1$. If leaders engage in behaviour that raises (lowers) their conspicuousness when flying towards (away from) the destination we would choose $\alpha_+ > 1$ ($\alpha_- < 1$). For bee swarms, streaking towards the nest by the scout leaders may serve to increase visibility, while ``laying low'' on return may decrease it \cite{Seeley}.

\subsection{100\% leader model}
\label{leader-only_model}

A leader-only model can be obtained by setting follower populations to zero ($u^{\pm}(x,t)=0$). As noted, attraction/repulsion social interactions are maintained, but the alignment bias is independent of the population. The target direction is potentially favoured through {\bf bias 1} ($\varepsilon$) and {\bf bias 2} ($\beta_+\ne\beta_-$, differential speeds). The model reduces to
\begin{eqnarray} \label{leader_only_model}
\pdifft{v^+} + \beta_+ \pdiffx{v^+} & = & -\lambda^{v^+} v^+ +\lambda^{v^-} v^-\,, \nonumber \\
\pdifft{v^-} - \beta_- \pdiffx{v^-} & = & +\lambda^{v^+} v^+ -\lambda^{v^-} v^-\,, \nonumber \\
v^\pm(x,0) & = & v_0^\pm(x)\,,
\end{eqnarray}
where 
\[
\lambda^{v^\pm} = \lambda_1 + \lambda_2 \left[ 0.5+0.5\tanh (y^{v^\pm}-y_0)\right], \, \textup{with } y^{v^\pm}= Q_{r}^{v^\pm}+Q_{a}^{v^\pm}+Q_{l}^{v^\pm}.
\]
The interaction contributions are given by
\begin{eqnarray} 
Q_{r}^{v^\pm} & = & q_r \int_{0}^{\infty} K_r(s) \left(v(x\pm s)-v(x \mp s) \right) ds\,, \\
Q_{a}^{v^\pm} & = & -q_a \int_{0}^{\infty} K_a(s) \left(v(x\pm s)-v(x \mp s) \right) ds\,, \\
Q_{l}^{v^\pm} & = & \mp 2 q_{l} \int_0^{\infty} K_{l}(s) \varepsilon ds  =  \mbox{constant} \label{Qal_M2} \,. 
\end{eqnarray}

\subsection{100\% follower model} 

We obtain a follower-only model by ignoring dynamic evolution of the leaders. Specifically, we stipulate fixed and uniform leader populations, i.e. $v^+(x,t)$ and $v^-(x,t)$ are constant in space and time. A leader contribution to attraction and repulsion is eliminated while their contribution to follower alignment is reduced to a fixed and constant bias, which we refer to as an
{\em implicit leader bias} and represent by parameter $\eta$: large $\eta$ corresponds to highly influential leaders. The resulting model is given by
\begin{eqnarray} \label{model_follower_only}
\pdifft{u^+} + \gamma \pdiffx{u^+} & = & -\lambda^{u^+} u^+ +\lambda^{u^-} u^-\,, \nonumber\\
\pdifft{u^-} - \gamma \pdiffx{u^-} & = & +\lambda^{u^+} u^+ -\lambda^{u^-} u^-\,, \nonumber \\
u^\pm(x,0) & = & u_0^\pm(x)\,,
\end{eqnarray}
where 
\[
\lambda^{u^\pm} = \lambda_1 + \lambda_2 \left[ 0.5+0.5\tanh (y^{u^\pm}-y_0)\right] \,, \, \textup{with } y^{u^\pm}= Q_{r}^{u^\pm}+Q_{a}^{u^\pm}+Q_{l}^{u^\pm}.
\]
and interaction terms 
\begin{eqnarray} 
Q_{r}^{u^\pm} & = & q_r \int_{0}^{\infty} K_r(s) \left(u(x\pm s)-u(x \mp s) \right) ds\,, \\
Q_{a}^{u^\pm} & = & -q_a \int_{0}^{\infty} K_a(s) \left(u(x\pm s)-u(x \mp s) \right) ds\,, \\
Q_{l}^{u^\pm} & = & q_l \int_0^{\infty} K_l(s) \left( u^{\mp} (x \pm  s) - u^{\pm} (x\mp s) \mp  \eta \right) ds  \label{Qal_M1} \,. \end{eqnarray}

\begin{table} 
\begin{center}
\caption{Summary of the interactions involved in the models. ``Full'' denotes Full follower-leader model;  ``LO'' denotes Leaders Only model; ``FO'' denotes Followers Only model.}
\label{table_interactions}       
\begin{tabular}{lllll} 
\multicolumn{1}{c}{\textbf{}} &\multicolumn{2}{c}{\textbf{Full}} & \multicolumn{1}{c}{\textbf{LO}} &  \multicolumn{1}{c}{\textbf{FO}} 
\\
 \noalign{\smallskip}  \hline
\\
\textbf{Pop. composition} &Leaders (L) &Followers (F) &Leaders (L) &Followers (F) 
\\
 \noalign{\smallskip}   \hline
\\
     \textbf{Attraction to} &F+L &F+L &F+L &F+L   
 \\
       \noalign{\smallskip}       \hline
\\
\textbf{Repulsion to} &F+L &F+L &F+L &F+L 
\\
          \noalign{\smallskip}    \hline 
\\
\textbf{Alignment to} &implicit      & F + L  &implicit &F + implicit      \\
 &orientation &(weighted &orientation &leader  \\
 &bias $\varepsilon$ &with $\alpha_\pm$) &bias $\varepsilon$ &bias $\eta$ \\
\noalign{\smallskip}\hline
\end{tabular}
\end{center}
\end{table}

\subsection{Parameters}

Given its complexity, the model has a large parameter set and we therefore fix many at standard values, based on previous studies \cite{Raluca} and listed in Appendix \ref{fixedparameters}. The fixed parameters include the follower speed $\gamma$ as well as the interaction ranges $s_r, s_l, s_a$, fixed to generate ``short-range repulsion, mid-range alignment and long-range attraction'', a common assumption in biological models of swarming behaviour \cite{Sumpter,Carrillo,fetecau2012mathematical}. Similarly, the more technical parameters $y_0, m_l, m_a, m_r$ are also chosen according to \cite{Raluca}, see Appendix  \ref{fixedparameters}.

Consequently, we focus on a smaller set of key parameters that distinguish leader/follower movement, listed in Table \ref{table1} along with the models to which they belong. In particular, we highlight the {\em bias} parameters that stipulate a level of attempted leader influence. We also remark that model formulations lead to conservation of follower and leader populations, generating two further population size parameters. As a final note, we generally restrict to alignment-attractive dominated regimes, i.e. $q_a, q_l \gg q_r$. 

\begin{table} 
\begin{center}
\caption{Table of parameters varied throughout this study. The parameters that are fixed throughout this study are summarised in Table \ref{table2} (see Appendix \ref{fixedparameters}). ``LO'' denotes Leaders Only model; ``FO'' denotes Followers Only model.}
\label{table1}       
\begin{tabular}{llll}
\multicolumn{1}{c}{\textbf{Grouping}} & \multicolumn{1}{c}{\textbf{Parameter}} & \multicolumn{1}{c}{\textbf{Description}} &  \multicolumn{1}{c}{\textbf{Model}}\\
    \hline
    \\
Bias: &$\alpha_+$            & alignment due to ($+$) oriented leaders & Full \\ 
                &$\alpha_-$            & alignment due to ($-$) oriented leaders& Full\\   
                &$\eta$         & implicit leader bias &FO \\ 
                &$\varepsilon$ &implicit orientation bias & LO, Full\\
                &$\beta_+$          &speed of ($+$) moving leaders &LO, Full    \\   
                &$\beta_-$         &speed of ($-$) moving leaders & LO, Full \\  
 \\
              \hline
\\
Pop. size: &$A_u$ &mean follower density   &FO, Full \\
                     &$A_v$ &mean leader density &LO, Full \\
                   &$M_u$ &maximum initial follower density &Full \\
                     &$M_v$ &maximum initial leader density &Full   \\
\\
              \hline
\\
Interaction: &$q_r$     &repulsion strength &All\\
&$q_l$     &alignment strength &All \\
&$q_a$     &attraction strength&All\\
\\
              \hline
\\
Others: &$\lambda_1$     &baseline turning rate &All\\
&$\lambda_2$     &bias turning rate &All\\
\noalign{\smallskip}\hline
\end{tabular}
\end{center}
\end{table}

\section{Dynamics of Leaders Only and Followers Only models}

We first analyse the dynamics of the simplified models, via linear stability analysis and numerical simulation. Note that details of the numerical scheme are provided in Appendix \ref{numericalmethods}.

\subsection{100\% leader model}

In this model, all swarm members have some knowledge of their target and bias their movement through two mechanisms: {\bf bias 1}, orientation according to the target and parametrised by $\varepsilon \ge 0$, and {\bf bias 2}, differential speed of movement, i.e. $\beta_+\ge\beta_-$.

\subsubsection{Steady states and stability analysis}
\label{analysis_leader_only}

\begin{figure}
    \centering
    {\includegraphics[width=1\textwidth]{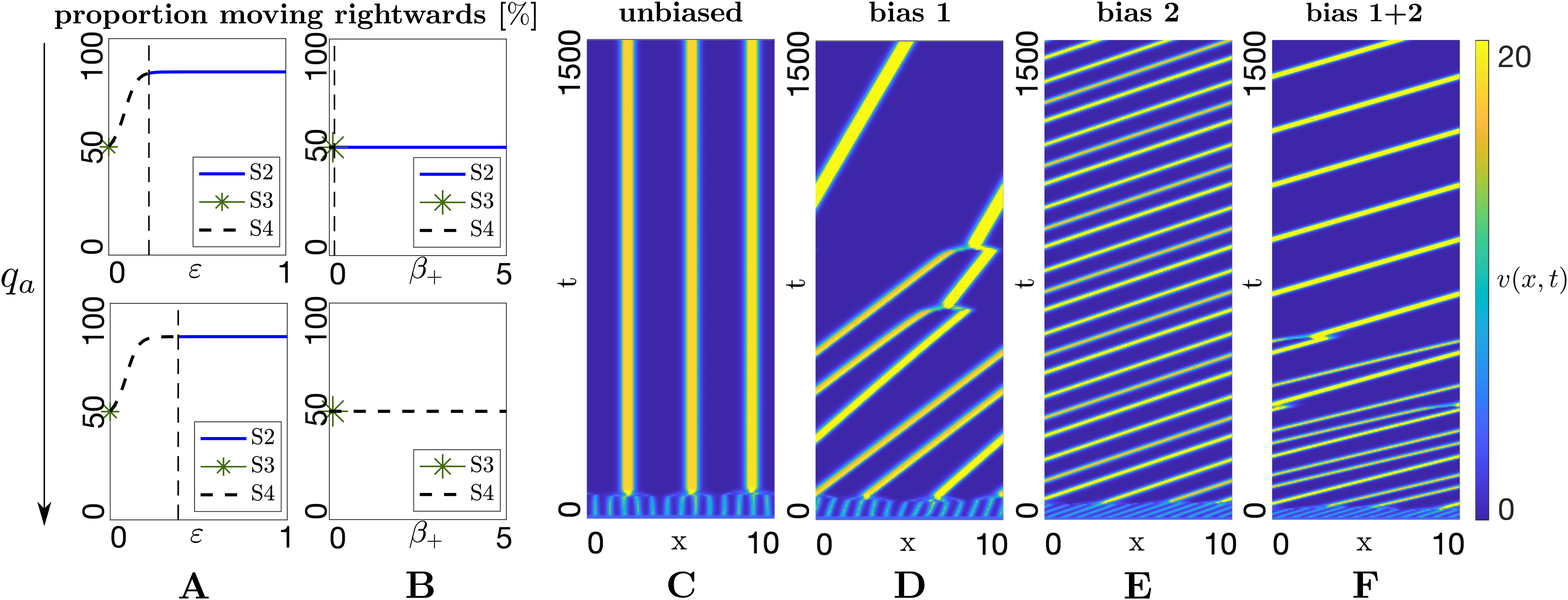}}
    \caption{Dynamics of the leader-only model. (A-B) Bifurcation plots showing HSS (proportion moving rightwards) and stability under parameter variation. (A) \textbf{Bias 1}, i.e. variation of $\varepsilon$ under low (top panel,  $q_a=0.5$) and high (bottom panel, $q_a=50$) attraction  (\textbf{bias 2} inactive, $\beta_+=\beta_-=0.1$). (B) \textbf{Bias 2}, varying $\beta_+$ for fixed $\beta_- = 0.1$, under low (top panel, $q_a=0.5$) and high (bottom panel, $q_a=50$) attraction (\textbf{bias 1} inactive, $\varepsilon=0$). Solid blue and dashed black lines denote stability class S2 and S4 respectively, dark green asterisks indicate stability class S3. (C-F) Space-time density map showing evolving total leader density, under: (C) unbiased case, generating a stationary patterns; (D) \textbf{bias 1}, obtained for $\varepsilon=0.2$ (\textbf{bias 2} inactive, $\beta_+=\beta_-=0.1$), generating target directed swarms;  (E) \textbf{bias 2}, obtained for $\beta_+ / \beta_- = 0.2 /0.1$ (\textbf{bias 1} inactive, $\varepsilon=0$), generating target directed swarms; (F) \textbf{biases 1 and 2}, obtained for $\varepsilon=0.2$ and $\beta_+ / \beta_- = 0.2 / 0.1$, generating target directed swarms with enhanced speed. (C-F) ICs are perturbations of $(v^*, v^{**})=(2,2)$. (D) ICs are perturbations of $(v^*, v^{**})=(3.329,0.671)$. In all plots, other parameters are set at $q_r=0.1$, $q_l=7.5$, ((C-F) $q_a=7.5$), $A_v=4,  \lambda_1=0.2, \lambda_2=0.9.$ 
    } \label{leader_only}
\end{figure}

We first examine the form and stability of spatially homogeneous steady state (HSS) solutions, $v^+(x,t)=v^*$ 
and $v^-(x,t)=v^{**}$, for the leader-only model (\ref{leader_only_model}-\ref{Qal_M2}). Conservation of mass leads to $A_v=v^*+v^{**}$, where $A_v$ is the sum of initial population densities averaged over space ($A_v = \left< v_0^+(x)+v_0^-(x) \right>$). The steady state equation is obtained by solving
\begin{equation} \label{hL=0}
h(v^*,q_l, \lambda, A_v, \varepsilon)=0,
\end{equation}
where 
\begin{eqnarray*}
h(v^*, q_l, \lambda, A_v, \varepsilon) & = & -v^*(1+\lambda \tanh(- 2\varepsilon q_l -y_0)) \\ & & + (A_v-v^*)(1+\lambda \tanh(2\varepsilon q_l -y_0))
\end{eqnarray*}
and
\begin{equation}
\lambda=\frac{0.5 \lambda_2}{0.5 \lambda_2+ \lambda_1}.
\end{equation}
From Eq. \ref{hL=0}, we obtain a single HSS solution
\begin{equation}\label{v^*}
v^*=\frac{A_v[1+\lambda \tanh(2 \varepsilon q_l-y_0)]}{2+\lambda \tanh(-2 \varepsilon q_l-y_0) + \lambda \tanh(2 \varepsilon q_l-y_0)}.    
\end{equation}
For $q_l=0$ (no alignment) or $\varepsilon=0$ (no {\bf bias 1}) we obtain an {\em unaligned} HSS $(v^*,v^{**})=\left( \frac{A_v}{2},\frac{A_v}{2} \right)$, i.e. a population equally distributed into those moving in ($\pm$) directions. Assuming $\varepsilon>0$, dominating alignment ($q_l \to \infty$) leads to steady state $(v^*,v^{**})=\left(A_v(1+\lambda)/2,A_v(1-\lambda)/2 \right)$. For $q_l>0$ the same result follows for dominating {\bf bias 1}, i.e. $\varepsilon \to \infty$. Intuitively, the introduction of bias eliminates symmetry, with $\varepsilon>0$ tipping the balance into a ($+$) direction, with alignment amplifying the effect. The steady state variation with $\varepsilon$ is illustrated in Figure \ref{leader_only}A. Unlike \textbf{bias 1}, introduction of differential leader speed does not alter the HSS solution, since $h$ does not depend on $\beta_\pm$, see Figure \ref{leader_only}B.

To assess stability and the potential for pattern formation we perform a standard linear stability analysis. Specifically, we examine the growth from homogeneous and inhomogeneous perturbations of the HSS at $(v^*,v^{**})=(v^*, A_v-v^*)$. Note that here it is convenient to extend $K_r$ and $K_a$  to odd kernels on the whole real line, i.e. 
\begin{eqnarray*}
\label{Qrep_Qattr_analysis}
Q_{r}^{v^\pm} & = & q_r \int_{-\infty}^{+\infty} K_r(s) v(x\pm s)  ds, \\
Q_{a}^{v^\pm} & = & -q_a \int_{-\infty}^{+\infty} K_a(s) v(x\pm s) ds. 
\end{eqnarray*}
We set $v^+(x,t)=v^*+v_p(x,t)$ and $v^-(x,t)=v^{**}+v_m(x,t)$, where $v_p(x,t)$ and $v_m(x,t)$ each denote small perturbations. We substitute into (\ref{leader_only_model}), neglect non-linear terms in $v_p$ and $v_m$ and look for solutions $v_{p,m} \propto e^{\sigma t + ikx}$. Here, $k$ is referred to the wavenumber (or spatial eigenvalue) while $\sigma$ is the growth rate  (or temporal eigenvalue). A few rearrangements lead to the expression
\begin{equation}
\label{disp_rel1}
\sigma^+ (k) = \frac{C(k)+\sqrt{C(k)^2 - D(k)}}{2},
\end{equation}
where $\sigma^+ (k)$ is used to denote the growth rate with largest real part. In the above
\begin{eqnarray*}
C(k) &=&
(\beta_- -\beta_+) i k
-2 \lambda_1 -\lambda_2 - 0.5 \lambda_2 \left[ \tanh(-2q_{l} \varepsilon-y_0) + \tanh(2q_{l}\varepsilon - y_0) \right], \\
D(k) &=& 4 \beta_+ \beta_- k^2 + 4i k \lambda_1 (\beta_+-\beta_-) \\
&& + 2 \lambda_2 i k  \{  \beta_+(1+ \tanh (2q_{l} \varepsilon-y_0))
- \beta_- (1+ \tanh (-2q_{l} \varepsilon-y_0)) \\ 
&&+v^* [ 1-\tanh^2 (-2 q_{l}\varepsilon-y_0) ][(-q_r \hat{K}_{r}^+(k) +q_a \hat{K}_{a}^+(k) ) (\beta_+ + \beta_-)]\\
&&+ v^{**} [ 1-\tanh^2 (2q_{l} \varepsilon-y_0) ][( q_r \hat{K}_{r}^-(k) -q_a \hat{K}_{a}^-(k))(\beta_+ +\beta_-)] \},
\end{eqnarray*}
where
$\hat{K}_{j}^\pm (k), j=r,a,l$ denote the Fourier transform of the kernel $K_j(s)$, i.e.
\[
\hat{K}_{j}^\pm (k)= \int_{-\infty}^{+\infty} K_j(s) e^{\pm iks} ds=\exp \left( \pm i s_j k-\frac{k^2 m_{l}^2}{2} \right), \quad j=r,a,l.
\]

The HSS is unstable (stable) to homogeneous perturbations if $\Re(\sigma^+(0))>0$ ($\Re(\sigma^+(0)) \leq 0$) and unstable to inhomogeneous perturbations if $\Re(\sigma^+(k))>0$ for at least one valid $k>0$ (for an infinite domain, we simply require $\Re(\sigma^+(k))>0$ for at least one value of $k\in \mathbb{R}^+$). Any $k$ for which $\Re(\sigma^+(k))>0$ is referred to as an unstable wavenumber. 

We classify HSS stability according to the following principle forms:
\begin{enumerate}
\item[(S1)] Unstable to homogeneous perturbations, i.e. $\Re(\sigma^+ (0)) > 0$. Solutions are expected to diverge from the HSS both with and without movement.
\item[(S2)] Stable to homogeneous and inhomogeneous perturbations, i.e. $\Re(\sigma^+ (k)) < 0, \quad \forall k \geq 0$. We expect small (homogeneous or inhomogeneous) perturbations to decay and solutions that evolve 
to the HSS.
\item[(S3)] Stationary patterns, HSS stable to homogeneous perturbations and unstable to inhomogeneous perturbations. Specifically, we have $\Re(\sigma^+(0)) \leq 0$ but $\exists \tilde{k} >0: \Re(\sigma^+(\tilde{k}))>0$ where, for any such $\tilde{k}$, $\Im(\sigma^+(\tilde{k}))=0$.
\item[(S4)] Dynamic patterns, as (S3), but $\Im(\sigma^+(\tilde{k})) \ne 0$ for at least some of the unstable wavenumbers. 
\end{enumerate}
(S3) and (S4) both indicate a Turing-type instability \cite{turing1952}, i.e. symmetry breaking in which a spatial pattern emerges from quasi-homogeneous initial conditions. The presence of wavenumbers where $\Im(\sigma^+(\tilde{k})) \ne 0$ implies growing patterns that oscillate in both space and time, potentially generating a dynamic pattern (e.g. a travelling swarm). These are, though, predictions based on solutions to the linearised system and nonlinear dynamics are likely to introduce further complexity.

Key results from the analysis are summarised in Figure \ref{leader_only}, indicating that both the HSS and its stability change with bias parameters $\varepsilon$ (or $q_l$), the ratio $\beta_+ / \beta_-$ and $q_a$. As noted above, increasing $\varepsilon$ (or $q_{l}$) generates a HSS with ($\pm$) distributions increasingly favouring the target direction. Variations in the $\beta_+ / \beta_-$ ratio do not alter the HSS value but do impact on the stability.
Under both biases 1 and 2, the stability nature changes at key threshold values, critically depending on the strength of attraction, $q_a$. For low $q_a$ the HSS is stable for all values of $\varepsilon$ and/or $\beta_+ / \beta_-$: attraction is insufficient to cluster the population and it remains dispersed. 
There may be biased movement towards the target, but the population remains in a uniformly dispersed/non-swarming state. 

For larger $q_a$, however, the HSS becomes unstable under inhomogeneous perturbations. A Turing-type instability occurs and emergence of a spatial pattern is expected. The predicted pattern critically depends on the bias. For an unbiased scenario ($\varepsilon=0$ and $\beta_+ / \beta_- =1$) we have stability class (S3) and predict a stationary pattern, see dark green asterisks in Figures \ref{leader_only}A and \ref{leader_only}B. Simulations corroborate this prediction (see Figure \ref{leader_only}C), where we observe stationary cluster formation. Each cluster is weighted equally between ($\pm$) directed populations and the overall cluster is fixed in position. Note, however, that the nonlocal elements of the model generate a degree of intercluster communication and, over longer timescales, clusters may attract each other and merge.

Introducing {\bf bias 1} ($\varepsilon > 0$) or {\bf bias 2} ($\beta_+ / \beta_- > 1$), though, generates growth rates with imaginary components -- this follows from the nonzero imaginary parts of $D(k)$ and/or $C(k)$ --
and the instability is of type (S4). In this case a dynamic component is predicted, with simulations substantiating this, cf. Figures \ref{leader_only}D and \ref{leader_only}E. The forming clusters are asymmetrically distributed between $(\pm)$ directed movement and, overall, we observe steered swarming: clusters move in the direction determined by the bias. Notably, clusters move at distinct speeds according to their size, so that clusters collide and merge. Eventually, a single steered swarm has formed and migrates with fixed speed and shape (a travelling pulse). The simultaneous action of biases 1 and 2 generates similar behaviour, Figure \ref{leader_only}F, with the combined action creating faster movement towards the target.

Summarising, the leader-only model illustrates the distinct contributions from different model elements: (i) attraction is crucial to aggregate a dispersed population; (ii) assuming sufficient attraction, either \textbf{bias 1} or \textbf{bias 2} is sufficient to propel the swarm in the direction of the target, with increased swarm speed if both biases act together.

\subsection{100\% follower model} 

We next examine the follower-only model. Interaction occurs through attraction, repulsion and alignment, with an additional
uniform alignment bias parametrised by $\eta$ and corresponding to implicit perception of a leader population.

\subsubsection{Steady states and stability analysis}

Proceeding as before, we explore the form of spatially homogeneous steady state solutions. Conservation of the total follower population leads to $A_u=u^*+u^{**}$, where $A_u$ is the average (over space) of the sum of initial population densities, $A_u = \left< u_0^+(x)+u_0^-(x) \right>$. Steady states will be given by
\begin{equation}
\label{h1=0}
h(u^*,q_{l}, \lambda, A_u, \eta)=0,
\end{equation}
where 
\begin{eqnarray*}
h(u^*, q_{l}, \lambda, A_u, \eta) & = & -u^*(1+\lambda \tanh(A_u q_{l}-2u^* q_{l} - \eta q_{l} -y_0)) \\
& & +(A_u-u^*)(1+\lambda \tanh(-A_u q_{l}+2u^* q_{l} + \eta q_{l} -y_0))
\end{eqnarray*}
and
\begin{equation}
\lambda=\frac{0.5 \lambda_2}{0.5 \lambda_2+ \lambda_1}.
\end{equation}

\begin{figure}
    \centering
    {\includegraphics[width=1\textwidth]{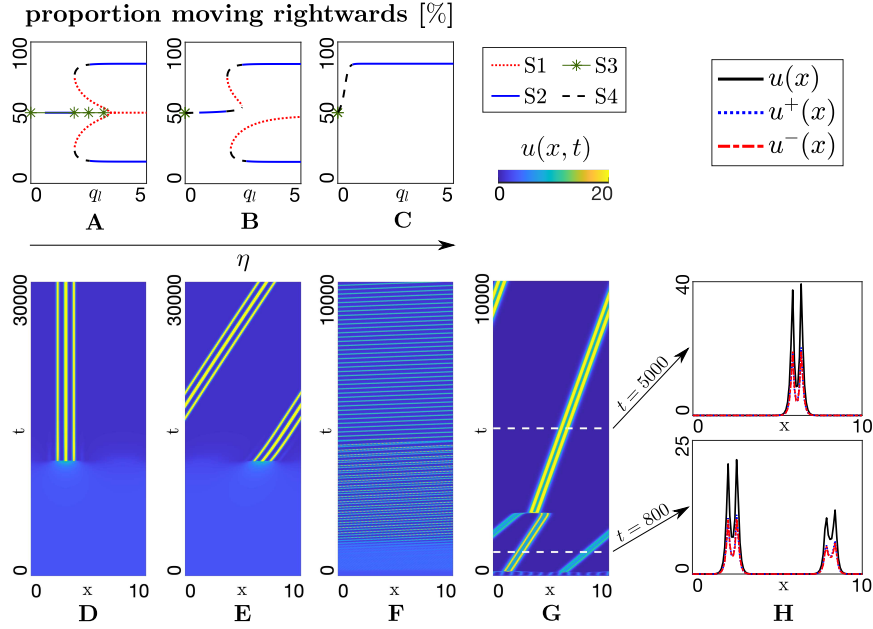}}
    \caption{Dynamics of the follower-only model. (A-C) Bifurcation plots showing HSS (proportion moving rightwards) and stability under parameter variation. Bifurcation parameter is $q_l$ and the resulting bifurcation plots are shown for (A) $\eta=0$ (unbiased), (B) $\eta=0.04$, (C) $\eta=4$. Other parameters set at $q_r=0, q_a=0.25, A_u=2, \lambda_1=0.8, \lambda_2=3.6$. Stability classes plotted as S1: dotted red, S2: solid blue, S3: dark green asterisks, S4: dashed black. (D-F) Space-time plot showing the evolving total follower density under variation of $\eta$: (D) $\eta=0$ (unbiased), (E) $\eta=0.04$, (F) $\eta=4$. Stronger biases lead to faster swarm movement towards the target. Other parameters set as in (A-C) with $q_l=0.4$. ICs are perturbations of $(u^*, u^{**})=(1,1)$. In (G-H) we demonstrate the merging of faster and slower swarms, under the parameter set $\eta=0.4, q_r=0.1, q_l=1, q_a=10,  A_u=2, \lambda_1=0.2, \lambda_2=0.9$.}
    \label{FO_model1}
\end{figure}

The zero-bias scenario ($\eta = 0$) has been analysed in depth previously, see \cite{Raluca}, and we restrict to a brief summary. First, a single {\em unaligned} HSS exists at
\[
(u^*,u^{**})=(\frac{A_u}{2},\frac{A_u}{2}),
\]
i.e. both directions equally favoured. Dominating alignment ($q_l \to \infty$) generates two further HSS at  $(u^{*},u^{**})=\left(A_u(1\mp\lambda)/2,A_u(1\pm\lambda)/2\right)$:  each {\em aligned} HSS corresponds to a population where alignment induces the population to favour one direction. A typical structure for the bifurcation diagram is illustrated in Figure \ref{FO_model1}A: a central branch corresponding to the unaligned HSS and upper and lower aligned branches. For the chosen parameters, these branches are connected via a further set of intermediate (unstable) branches. Thus, as $q_{l}$ increases the number of steady states shifts between 1, 5 and 3 steady states (see also Figure \ref{135_ss_qal} of  Appendix \ref{follower_only_steadystates}). 

The symmetric structure of $\eta = 0$ is lost for $\eta \ne 0$, even under small values: see Figures \ref{FO_model1}B-C. The aligned HSS branch corresponding to the target direction is more likely to be selected, the other branch is shifted rightwards (Figure \ref{FO_model1}B) and for larger $\eta$ disappears entirely (Figure \ref{FO_model1}C). Overall, the external bias is amplified by follower to follower alignment and the population becomes predominantly oriented in the target direction.

We extend to a spatial linear stability analysis, applying the same process as in section \ref{analysis_leader_only} to obtain the following dispersion relation
\begin{equation}
\sigma^+ (k) = \frac{C(k)+\sqrt{C(k)^2 - D(k)}}{2},
\label{disp_rel}
\end{equation}
where
\begin{eqnarray}
\label{C}
C(k) &= & \left( \lambda^{u^+}_{u^-} -\lambda^{u^+}_{u^+} \right) u^+ + \left( \lambda^{u^-}_{u^+} - \lambda^{u^-}_{u^-} \right) u^- -\lambda^{u^-} -\lambda^{u^+}, \\
D(k) &= & 4 \gamma^2 k^2 \nonumber \\ 
& & + 4\gamma i k \left[ \left( -\lambda^{u^+}_{u^-}-\lambda^{u^+}_{u^+} \right)u^+ + \left( \lambda^{u^-}_{u^-}+\lambda^{u^-}_{u^+} \right) u^- + \lambda^{u^-} - \lambda^{u^+} \right].\label{D}
\end{eqnarray}
In the above, $\lambda^{u^\pm}_{u^{\pm}}$ denote the partial derivatives of $\lambda^{u^\pm}$ with respect to $u^{\pm}$ and subsequently evaluated at the HSS $(u^*,u^{**})$.  For reference we provide the explicit forms in Appendix \ref{follower_only_lin_stab_appendix}, yet intricacy of the dispersion relation restricts us to a numerical approach. Stability is again classified into one of the 4 classes described earlier. 

The diagrams shown in Figure \ref{FO_model1}A-C reveal a complex bifurcation structure and potentially diverse dynamics according to parameter selection and initial condition. Indeed, this has already been highlighted in depth for the unbiased ($\eta=0$) model in \cite{Raluca}, where various complex spatiotemporal pattern forms have been revealed. For example, Figure \ref{simu_bif_qal_follower_only} in Appendix \ref{alignment_equilibrium_points} illustrate transitioning between stationary and dynamic aggregates as the key parameter $q_l$ is altered. Note that moving aggregates can be generated without any incorporated bias, though if the population begins quasi-symmetric either direction will be selected with equal likelihood.

Here we focus on the extent to which introduction of a bias influences the dynamics of aggregate structures, with Figure \ref{FO_model1}D-G providing a representative sequence. We begin with an unbiased scenario, setting $\eta=0$ and choosing parameters from a region predicted to lead to stationary patterning. We initiate populations in quasi-symmetric fashion, setting 
\[
\begin{array}{ll} u^+(x,0) = \dfrac{A_u \left(1+r_u(x)\right)}{2}\,, \quad & u^-(x,0) = \dfrac{A_u \left(1-r_u(x)\right)}{2}\,, \end{array}
\]
where $r_u(x)$ denotes a small random perturbation. As expected from the stability analysis, a stationary cluster forms (see Figure \ref{FO_model1}D) with its shape and position maintained by a symmetric distribution of $(\pm)$ directed populations. Introducing bias, though, disrupts the symmetry and Turing instabilities falls into the dynamic pattern class. Moreover, even a marginal alignment bias strongly selects clusters that move in the direction of the bias, e.g. see Figures \ref{FO_model1}E. Starting from a symmetric or nonaligned initial set-up, bias slightly tilts followers towards the target. Follower-follower alignment snowballs, eventually resulting in a cluster moving towards the target. Increasing the bias magnitude increases swarm speed, Figure \ref{FO_model1}F.

As for the 100\% leader model, there is a clear relationship between cluster speed and cluster size. This is illustrated in Figure \ref{FO_model1}G, where the initial symmetry breaking process generates two clusters of slightly different size, Figure \ref{FO_model1}H(bottom). Both clusters move in the target direction, but the smaller cluster is considerably faster. The clusters eventually collide and merge to form an even larger and slower cluster, see Figure \ref{FO_model1}H(top). Note that, in principle it is also possible to obtain a swarm migrating oppose the target direction, e.g. by heavily favouring biasing the initial conditions. Simulations, though, suggest that such situations are highly unlikely to occur in practice.



\begin{figure}
    \centering
    {\includegraphics[width=1\textwidth]{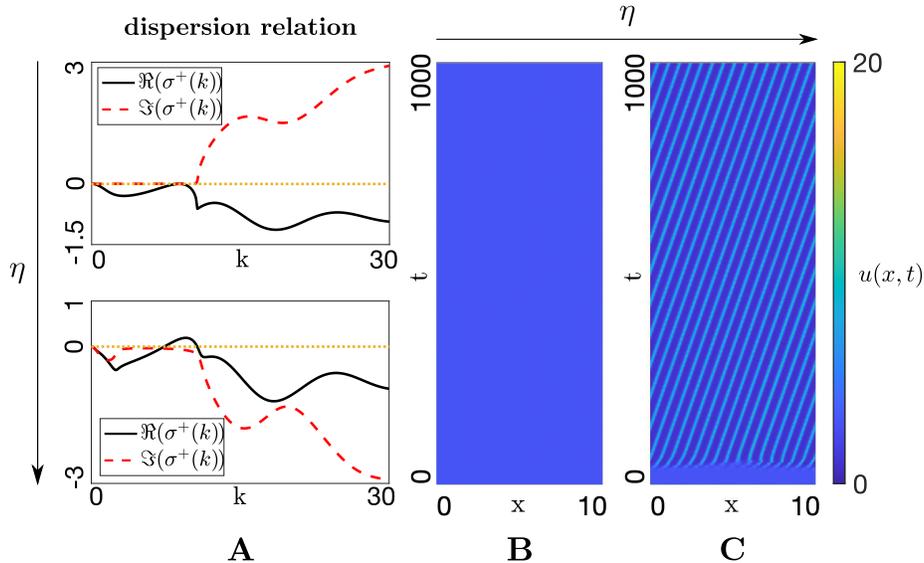}}
    \caption{(A) Dispersion relations and (B-C) the corresponding numerical simulations for the following parameter values:  $q_r=0, q_l=2.1, q_a=0, A_u=2, \lambda_1=0.8, \lambda_2=3.6$. (A) top row and (B) assume no bias ($\eta=0$), while (A) bottom row and (C) consider the effect of a small alignment bias ($\eta=0.08$). (B-C) ICs are perturbations of ($u^*, u^{**}$) = (1,1).}
    \label{symmetry_breaking}
\end{figure}

Introducing bias can even trigger symmetry breaking, as shown in Figure \ref{symmetry_breaking}. To highlight this, we neglect attractive and repulsive interactions ($q_a = q_r = 0$) and focus solely on alignment. Initially setting $\eta = 0$, remaining parameters are specified such that the unaligned HSS (i.e. $u^* = u^{**}$) is stable to both homogeneous and inhomogeneous perturbations: a typical dispersion relation is provided in Figure \ref{symmetry_breaking}A (top), showing the absence of wavenumbers with positive growth rates and the corresponding simulation confirms the absence of pattern formation, Figure \ref{symmetry_breaking}B. Introducing bias ($\eta > 0$) breaks symmetry, yielding a nonzero range of wavenumbers with positive growth rates, Figure \ref{symmetry_breaking}A (bottom). A pattern emerges which generates multiple clusters moving in the target direction, Figure \ref{symmetry_breaking}C.

Summarising the analysis and numerics in this section, we emphasize that the follower only submodel can display a range of aggregating/swarming behaviour, where the processes of alignment and attraction combine to generate one or more cluster. The addition of bias breaks directional symmetry, eliminating the formation of stationary structures and generating clusters that move coherently in the target direction.

\section{Dynamics of the full follower-leader model}

We turn attention to the full ``follower-leader'' model, formed from Equations (\ref{model}-\ref{Qlu_full_model}) and where followers constitute a completely naive population. Our principal aim will be to understand whether a steered swarm can arise under leader generated bias. Relying principally on numerical simulation, we focus on two general parameter regimes: (P1) strong attraction-strong alignment, and (P2) strong attraction-weak alignment. For simplicity we neglect repelling interactions ($q_r = 0$). A bias corresponding to the ($+$)-direction can occur through parameter choices:
\begin{itemize}
    \item {\bf Bias 1}, $\varepsilon > 0$, orientation;
    \item {\bf Bias 2}, $\beta_+ > \beta_-$, speed;
    \item {\bf Bias 3}, $\alpha_+ > \alpha_-$, conspicuousness.
\end{itemize}
Consequently, the {\em unbiased case} is $\varepsilon = 0$, $\alpha_+ = \alpha_-$, and $\beta_+ = \beta_-$. As discussed earlier, evidence is found for each bias in our honeybee swarming exemplar. Note that parameter regimes are selected such that linear stability analysis of the uniform solution in the unbiased case predicts Turing pattern formation. 

\subsection{Steady state analysis}

Steady state analysis proceeds as before: we look for the spatially homogeneous steady states $u^+(x,t)=u^*$, $u^-(x,t)=u^{**}$ and $v^+(x,t)=v^*$, $v^-(x,t)=v^{**}$, noting that conservation ensures $A_u=u^*+u^{**}$ and $A_v=v^*+v^{**}$, where $A_u$ and $A_v$ are as earlier described. Steady states for the full model satisfy
\begin{eqnarray}
\label{h=0}
h_u(u^*,q_l, \lambda, A_u, A_v, \alpha_-, \alpha_+, y_0)=0, \\
h_v(v^*,q_l, \lambda, A_v, \varepsilon, y_0)=0,
\end{eqnarray}
where 
\begin{eqnarray*}
h_u & =&-u^*(1+\lambda \tanh(A_u q_l-2u^* q_l + q_l \alpha_-(A_v-v^*)-q_l \alpha_+ v^* -y_0)) \\
&  &+(A_u-u^*)(1+\lambda \tanh(-A_u q_l+2u^* q_l + q_l \alpha_+ v^*-q_l \alpha_-(A_v- v^*) -y_0)), \\
h_v & =&-v^*(1+\lambda \tanh(-2 \varepsilon q_l-y_0))+(A_v-v^*)(1+\lambda \tanh(2 \varepsilon q_l- y_0)),
\end{eqnarray*}
and
\begin{equation}
\lambda=\frac{0.5 \lambda_2}{0.5 \lambda_2+ \lambda_1}.
\end{equation}
Leader steady states correspond to those obtained previously for the leader-only model. Hence, the proportion of leaders at HSS moving in the ($+$) direction increases monotonically between $A_v/2$ and $A_v(1+\lambda)/2$, according to $\varepsilon$ and/or $q_l$, (Figure \ref{leader_only}A). This equivalence stems from the simplification that leaders ignore others with respect to alignment.

In absence of alignment, i.e. $q_l=0$, we find a single unaligned HSS at $(u^*, u^{**},v^*,v^{**})=(A_u/2,A_u/2,A_v/2,A_v/2)$.
If $q_l \neq 0$, follower steady states are clearly more complex and we first consider the unbiased case ($\varepsilon = 0$, $\alpha_+ = \alpha_-$, $\beta_+ = \beta_-$). Here we have $v^*=v^{**}=A_v/2$ and hence
\begin{eqnarray*}
h_u & =&-u^*(1+\lambda \tanh(A_u q_l-2u^* q_l - y_0)) \\
& &+(A_u-u^*)(1+\lambda \tanh(-A_u q_l+2u^* q_l - y_0))\,.
\end{eqnarray*}
Leaders have no influence and follower steady states are as observed for the follower-only model with $\eta = 0$. As described earlier, the number of follower steady states varies between 1, 3 and 5 (see Figure \ref{135_ss_qal} of the Appendix \ref{follower_only_steadystates}) with sufficiently large alignment allowing followers to self-organise into a dominating orientation. 

We next consider an extreme {\bf bias 1} ($\varepsilon \to \infty$), while
excluding other biases ($\alpha=\alpha_+=\alpha_-, \beta_+ = \beta_-$). Leaders favour the ($+$) direction, specifically $(v^*,v^{**})=(\frac{A_v(1+\lambda)}{2}, \frac{A_v(1-\lambda)}{2})$, and hence
\begin{eqnarray*}
h_u & = & -u^*(1+\lambda \tanh(A_u q_l-2u^* q_l - q_l \alpha A_v \lambda - y_0)) \\
& & + (A_u-u^*)(1+\lambda \tanh(-A_u q_l + 2 u^* q_l + q_l \alpha A_v \lambda -y_0))\,.
\end{eqnarray*}
The above has the same structure as for the follower-only model under external bias, where $\eta$ in Equation \eqref{h1=0} is replaced by $\alpha A_v \lambda$. Consequently, for either increasing leader to follower influence ($\alpha$) or increasing leader population size ($A_v$), bifurcations occur as in Figure \ref{FO_model1}A-C: symmetric follower steady states become asymmetric, favoured according to the bias. 

\begin{figure}
    \centering
    {\includegraphics[width=0.8\textwidth]{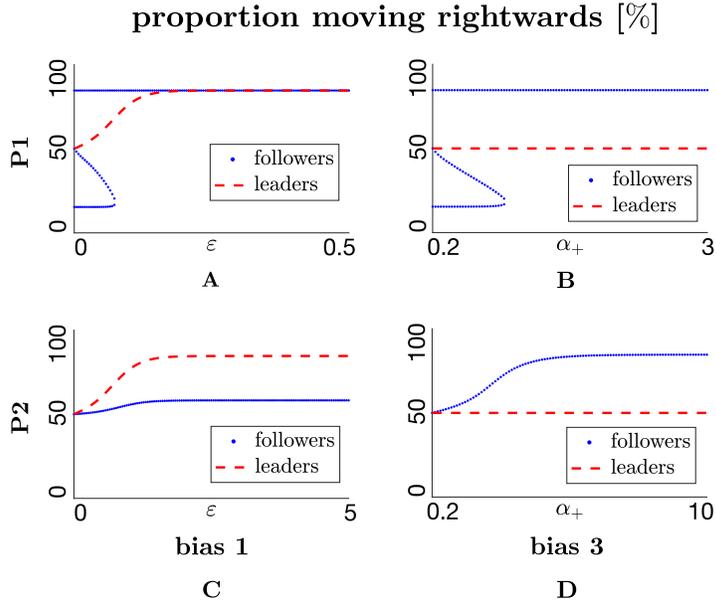}}
    \caption{Proportion of right-moving populations at steady state(s). (A,C) Effect of \textbf{bias 1}, increasing $\varepsilon$, on position and number of equilibrium points (\textbf{bias 2 and 3} inactive, $\beta_-=\beta_+=0.1, \alpha_-=\alpha_+=1$). (B,D) Effect of \textbf{bias 3}, increasing $\alpha_+ / \alpha_-$, on position and number of equilibrium points, for $\alpha_-=0.2$ (\textbf{bias 1 and 2} inactive, $\varepsilon=0$ and $\beta_-=\beta_+=0.1$). Top row corresponds to (P1) strong attraction-strong alignment ($q_r=0, q_l=10, q_a=8$), bottom row corresponds to (P2) strong attraction-weak alignment ($q_r=0, q_l=1, q_a=10$). Other parameter values fixed at $A_u=A_v=1$, $\lambda_1=0.2, \lambda_2=0.9$.}
\label{M3_epsilon_alpha}
\end{figure}

Differential speeds (bias 2, $\beta_+\ne \beta_-$) do not impact on steady states and we turn instead to distinct conspicuousness, specifically extreme {\bf bias 3} ($\alpha_+/\alpha_- \rightarrow \infty$) while eliminating {\bf bias 1}. Leader steady states remain symmetrical  ($v^* = v^{**} = A_v/2$), yet distinct conspicuousness tips the majority of followers to the bias direction and a single steady state occurs at
\[(u^*, u^{**}, v^*, v^{**})=(A_u (1+\lambda)/2, A_u (1-\lambda)/2, A_v/2, A_v/2)\,.\]
The bifurcation diagrams in Figure \ref{M3_epsilon_alpha} numerically confirm these results. Finally, we note that as $q_l \to \infty$ two further HSS's arise at $(u^*, u^{**}, v^*, v^{**})=(A_u (1 \mp \lambda)/2, A_u (1 \pm \lambda)/2, A_v (1 + \lambda)/2, A_v (1 - \lambda)/2)$.

\subsection{Numerical simulation}

The steady state analysis provides insight into whether different biases induce left-right asymmetry, yet the emerging dynamics of spatial structures remains unclear. We numerically explore the full spatial nonlinear problem, in particular its capacity to generate a steered swarm as described earlier. Simulations will be conducted for two forms of initial condition.
\begin{enumerate}
\item[(IC1)] {\em Unbiased and dispersed}. Populations quasi-uniformly distributed in space and orientation. Letting $A_u$ and $A_v$, respectively denote the mean total follower and leader densities, 
\begin{eqnarray*}
u^+(x,0) = \dfrac{A_u \left(1+r_u(x)\right)}{2}\,, & \quad & u^-(x,0) = \dfrac{A_u \left(1-r_u(x)\right)}{2}\,, \\
v^+(x,0) = \dfrac{A_v\left(1+r_v(x)\right)}{2}\,, & \quad & v^-(x,0) = \dfrac{A_v\left(1-r_v(x)\right)}{2}\,.
\end{eqnarray*}
\item[(IC2)] {\em Unbiased and aggregated}. Populations initially aggregated but unbiased in orientation. Letting $M_u$ and $M_v$ respectively denote the maximum initial follower and leader densities, 
\begin{eqnarray*}
u^+(x,0) = \dfrac{M_u e^{-5(x-x_0)^2} \left(1+r_u(x)\right) }{2}\,,
& \quad & u^-(x,0) = \dfrac{M_u e^{-5(x-x_0)^2} \left(1-r_u(x)\right)}{2}\,, \\
v^+(x,0) = \dfrac{M_v e^{-5(x-x_0)^2} \left(1+r_v(x)\right) }{2}\,,
& \quad & v^-(x,0) = \dfrac{M_v e^{-5(x-x_0)^2} \left(1-r_v(x)\right)}{2}\,.
\end{eqnarray*}
\end{enumerate}
Note that $r_u(x), r_v(x)$ denote small (1\%) random perturbations. (IC1) allow investigation into whether dispersed populations self-organise into swarms while (IC2) tests whether aggregated populations maintain a swarm profile. (IC2) are particularly appropriate for bee swarming, where followers and leader scouts are initially clustered together.

\begin{figure}[t!]
{\includegraphics[width=\textwidth]{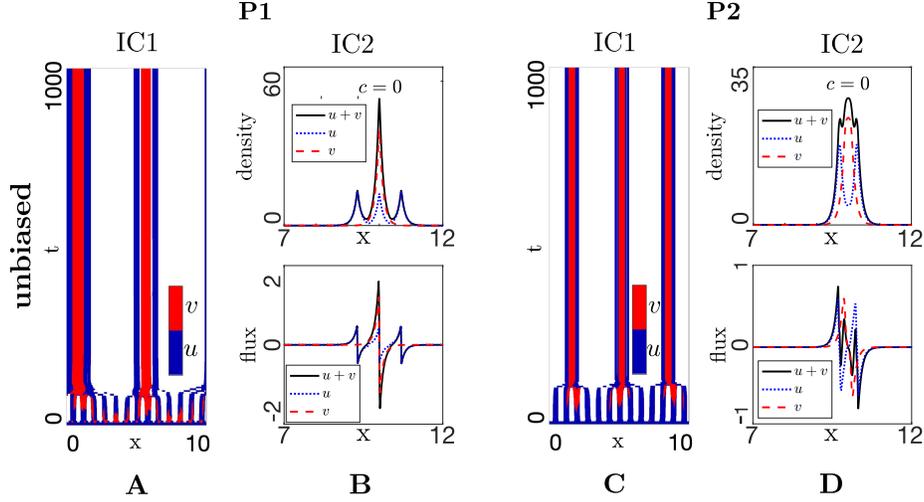}}
\caption{Dynamics of the full-model, unbiased case, obtained for $\varepsilon=0, \beta_-=\beta_+=0.1, \alpha_-=\alpha_+=1$. (A,C) Space-time evolution of densities under (IC1) for (A) P1, strong attraction-strong alignment, (C) P2, strong attraction-weak alignment. Non-white regions indicate where the total population $> 2(A_u+A_v$), i.e. ``clustering'' has occurred, with red/blue indicating local predominance of leaders/followers respectively. Panels (B,D) show (top) population distribution and (bottom) population fluxes for solutions under (IC2) for (B) P1, strong attraction-strong alignment, (D) P2, strong attraction-weak alignment. (P1) $q_r = 0, q_l = 10, q_a = 8$, (P2) $q_r = 0, q_l = 1, q_a = 10$, with other parameter value set as $A_u=A_v=1$ (IC1), $M_u=M_v=12.61$ (IC2), $\lambda_1=0.2, \lambda_2=0.9$.}
\label{unbiased}
\end{figure}

\subsubsection{Unbiased dynamics}

We first explore the capacity for self-organisation in the unbiased scenario. Note that each of the two principal parameter sets were selected to generate Turing instabilities and Figure \ref{unbiased}A and C demonstrate the patterning process under (P1) strong attraction-strong alignment and (P2) strong attraction-weak alignment, respectively. We observe the formation of multiple swarms which, in the absence of bias, remain in more or less fixed positions. Note, though, that over longer timescales inter-aggregate interactions may lead to drifting and merging. The arrangement and behaviour of an isolated swarm is investigated by initially aggregating the populations as in (IC2), with reorganisation leading to a stable and stationary swarm configuration and computed swarm wavespeed $c = 0$, Figure \ref{unbiased}B and D. Swarms contain leaders\footnote{These are leaders in name only, as in the unbiased scenario there is no directional bias in force.} concentrated at the swarm centre, with followers symmetrically dispersed either side. The distinct follower/leader profiles arise as leaders only interact through attraction, while followers receive additional alignment information. We further plot the {\em fluxes}, i.e. the quantities $u^+(x,t)-u^-(x,t)$ and $v^+(x,t)-v^-(x,t)$. In the stationary swarm profile, ($\pm$) movement is balanced such that the swarm maintains its position and shape, see Figure \ref{unbiased}.

\subsubsection{Introduction of leader biases}

\begin{figure}[t!]
{\includegraphics[width=\textwidth]{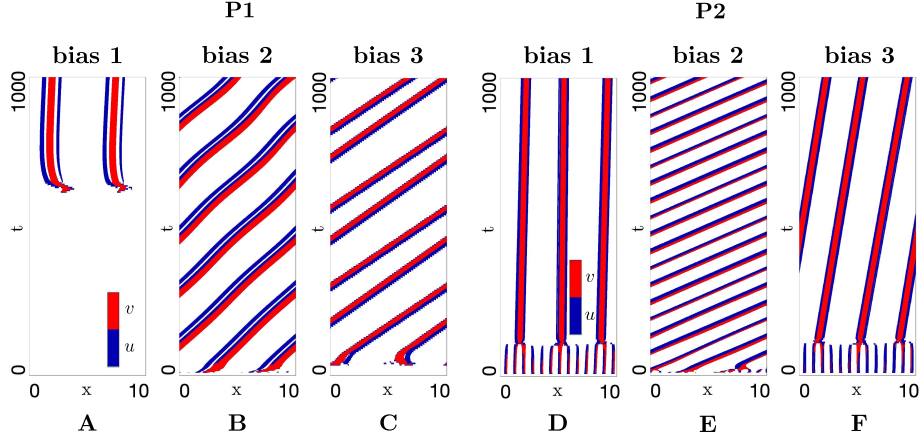}}
\caption{Impact of biases on swarm movement for the full model in a strong attraction-strong alignment regime (P1), strong attraction-weak alignment regime (P2) and under (IC1). Populations plotted in space-time (colourmap as described in Figure \ref{unbiased}). (A,D) \textbf{Bias 1}, obtained for $\varepsilon=0.2$ (\textbf{bias 2 and 3} inactive, $\beta_-=\beta_+=0.1, \alpha_-=\alpha_+=1$); 
(B,E) \textbf{bias 2}, obtained for $\beta_+ / \beta_-=0.5 / 0.1$ (\textbf{bias 1 and 3} inactive, $\varepsilon=0, \alpha_-=\alpha_+=1$); (C,F) \textbf{bias 3}, obtained for $\alpha_+ / \alpha_-=1.0 / 0.2$ (\textbf{bias 1 and 2} inactive, $\varepsilon=0, \beta_-=\beta_+=0.1$). Remaining parameters set at
(P1) $q_r=0, q_l=10, q_a=8$, (P2) $q_r=0, q_l=1, q_a=10$ and $A_u=A_v=1$, $\lambda_1=0.2, \lambda_2=0.9$. 
}
\label{Uniform_IC_bias}
\end{figure}


We perform the same set of simulations, but extended to include one of the three proposed mechanisms for leader bias. Simulation results under (IC1) indicate that self-organisation can be maintained under the inclusion of leader-bias, where again we observe that an initially dispersed population aggregates into one or more swarm, see Figure \ref{Uniform_IC_bias}. Notably, these swarms can subsequently migrate through space, indicating that a leader-generated bias can lead to sustained swarm movement. Yet the degree and direction of movement significantly varies with the type (and strength) of bias, demanding a more extensive analysis of when and which type of bias leads to steered swarm movement.

\begin{figure}[t!]
\begin{center}
\includegraphics[width=\textwidth]{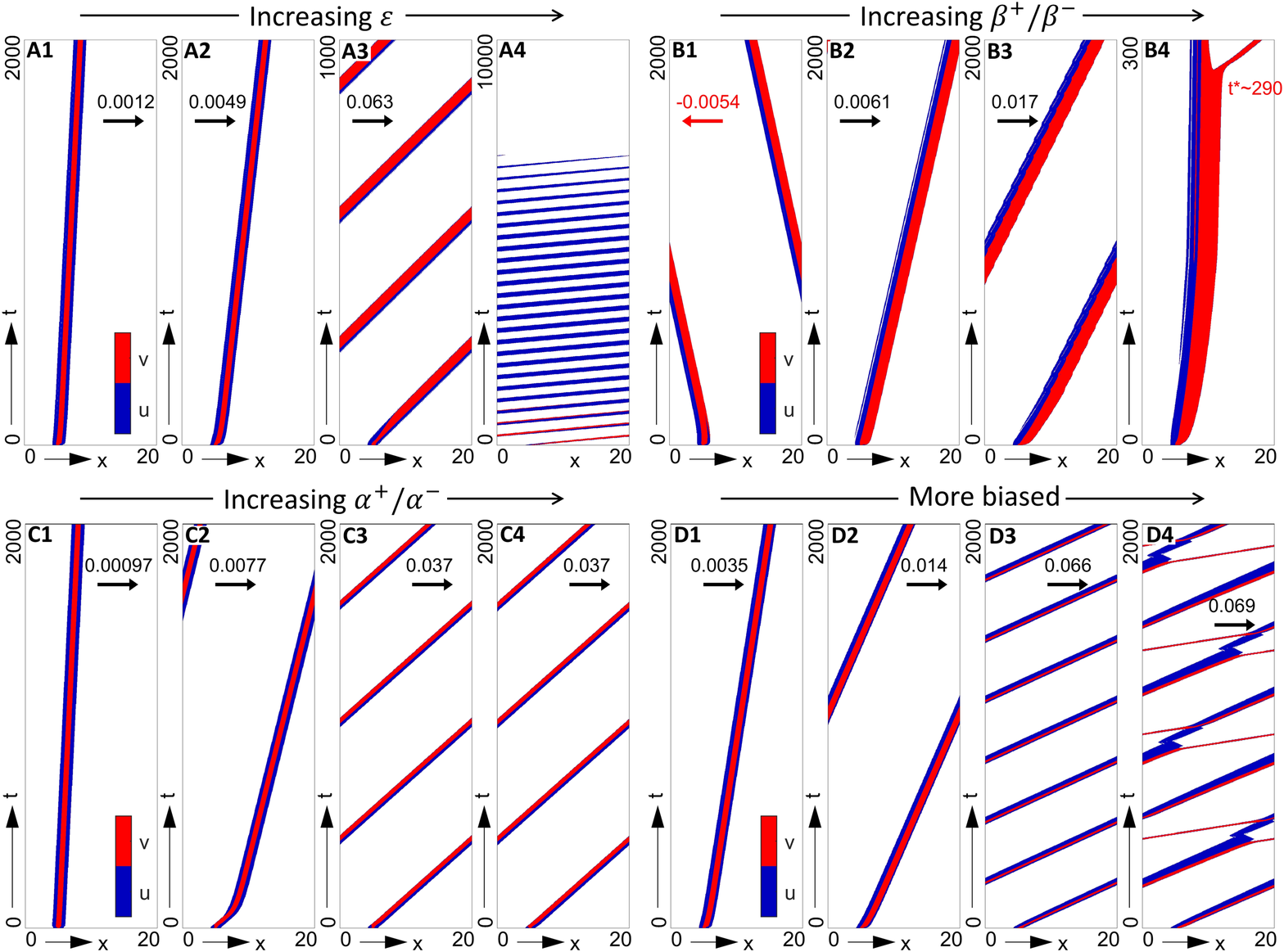}
\end{center}
\caption{Impact of biases on swarm movement for the full model in a strong alignment-strong attraction regime (P1) and under (IC2). Populations plotted in space-time (colourmap as described in Figure \ref{unbiased}). Note that we append each plot with the swarm speed, for cases where a travelling wave solution is (numerically) found. (A) {\bf Bias 1}, obtained for $\varepsilon = $ (A1) 0.5, (A2) 1.0, (A3) 2.0, (A4) 3.0 (bias 2 and 3 inactive, $\beta_+=\beta_-=0.1,  \alpha_+=\alpha_-=1$). (B) {\bf Bias 2}, obtained for $\beta_+/\beta_- = $ (B1) 0.2/0.1, (B2) 0.3/0.1, (B3) 0.5/0.1, (B4) 0.6/0.1 (bias 1 and 3 inactive, $\varepsilon = 0, \alpha_+=\alpha_-=1$). (C) {\bf Bias 3}, obtained for $\alpha_+/\alpha_- = $ (C1) 1/0.9, (C2) 1/0.575, (C3) 1/0.55, (C4) 1/0  (bias 1 and 2 inactive, $\varepsilon = 0, \beta_+=\beta_-=0.1$). (D) Simultaneous biases, for (D1) $\varepsilon = 0.25, \beta_+/\beta_-=0.15/0.1, \alpha_+/\alpha_-=1/2/3$,
(D2) $\varepsilon = 0.5, \beta_+/\beta_-=0.2/0.1, \alpha_+/\alpha_-=1/0.5$,
(D3) $\varepsilon = 0.75, \beta_+/\beta_-=0.25/0.1, \alpha_+/\alpha_-=1/0.4$,
(D4) $\varepsilon = 1, \beta_+/\beta_-=0.3/0.1, \alpha_+/\alpha_-=1/(1/3)$.
Other paramenters are (P1) $q_r = 0, q_l = 10, q_a = 8$, $M_u=M_v=12.61$, $\lambda_1 = 0.2, \lambda_2 = 0.9$.}\label{BiasesSetA}
\end{figure}

\begin{figure}[t!]
\begin{center}
\includegraphics[width=\textwidth]{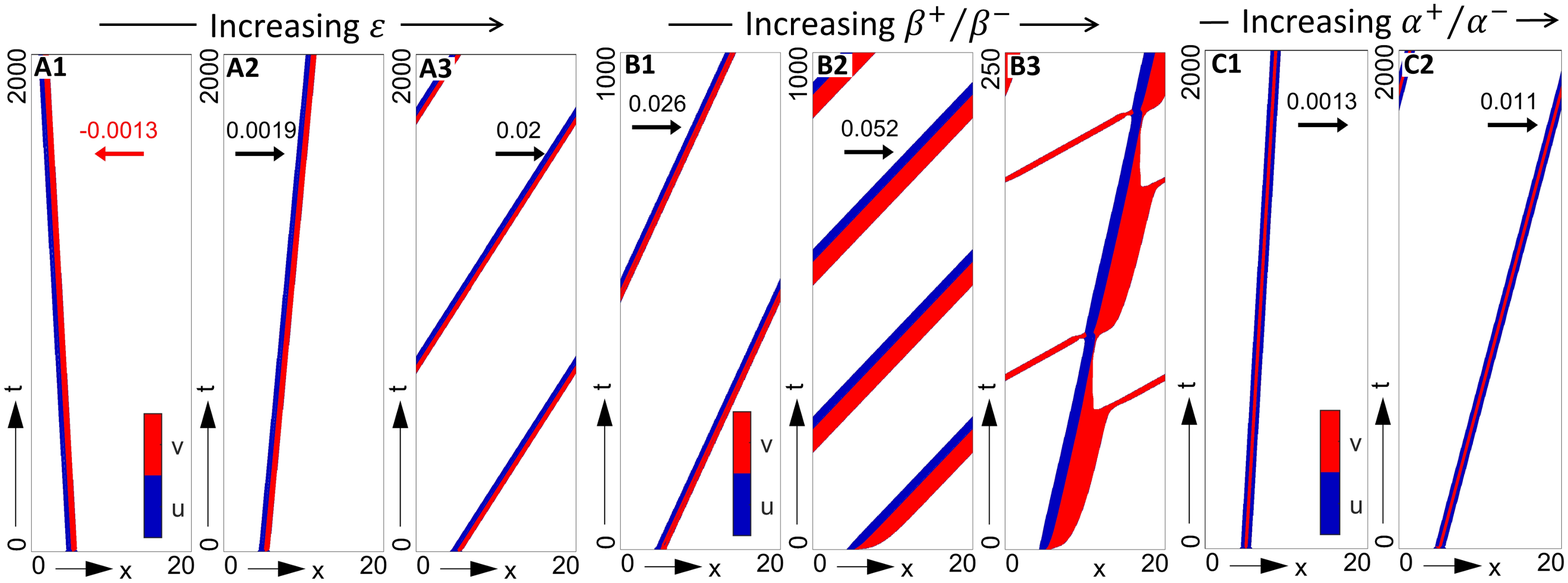}
\end{center}
\caption{Impact of biases on swarm movement for the full model in a strong attraction-weak alignment regime (P2) and under (IC2). Populations plotted in space-time (colourmap as described in Figure \ref{unbiased}). Note that we append each plot with the swarm speed, for cases where a travelling wave solution is (numerically) found. (A) {\bf Bias 1}, obtained for $\varepsilon = $ (A1) 3.0, (A2) 5.0, (A3) 15.0 (\textbf{bias 2 and 3} inactive, $\beta_+=\beta_-=0.1, \alpha_+=\alpha_-=1$). (B) {\bf Bias 2}, obtained for $\beta_+/\beta_- = $
(B1) 0.2/0.1, (B2) 0.8/0.1, (B3) 1/0.1 (\textbf{bias 1 and 3} inactive, $\varepsilon = 0, \alpha_+=\alpha_-=1$). (C) {\bf Bias 3}, obtained for $\alpha_+/\alpha_- = $ (C1) 1/0.9, (C2) 1/0 (\textbf{bias 1 and 2} inactive, $\varepsilon = 0, \beta_+=\beta_-=0.1$). Other parameters are (P2) $q_r = 0, q_l = 10, q_a = 8$, $M_u=M_v=12.61$, $\lambda_1 = 0.2, \lambda_2 = 0.9$.}\label{BiasesSetC}
\end{figure}

To investigate this in a controlled manner, we force populations into forming an isolated swarm by applying (IC2), ensuring that any subsequent swarm dynamics are the result of internal interactions rather than the influence of neighbouring swarm profiles. Each of the bias strengths are then progressively altered, individually or in concert, under each of our two principal parameter regimes (strong attraction-strong alignment and strong attraction-weak alignment): Figures \ref{BiasesSetA} and  \ref{BiasesSetC}
respectively plot the key behaviours observed for these two regimes.

The dynamics generated by {\bf bias 1} are illustrated in Figure \ref{BiasesSetA}A and Figure \ref{BiasesSetC}A. Over a wide range of bias strengths, {\bf bias 1} generates steered swarming, with an increased speed in the target direction as $\varepsilon$ increases. However two caveats must be highlighted. First, under certain parameter combinations we unexpectedly observe swarms that move away from the the target, specifically for weaker biases in the weak alignment regime (see Figure \ref{BiasesSetC}A1). Second, excessive biases can lead to loss of swarm coherence and eventual dispersion (see Figure \ref{BiasesSetA}A4). Thus, we conclude {\bf bias 1} is found to be only partially successful in generating a steered swarm.

We next consider {\bf bias 2}, i.e. increasing the ratio of leader speeds when moving towards or away from the target. Indicative simulations are plotted in Figures \ref{BiasesSetA}B and Figure \ref{BiasesSetC}B. Similar to {\bf bias 1}, successful steering only occurs within a range of $\beta_+/\beta_-$ values. First, as observed above under {\bf bias 1}, certain parameter regimes are capable of generating counter target directed swarms, for example see Figure \ref{BiasesSetA}B1 for a moderately faster $v^+$ population in the high attraction-high alignment regime. Second,
while increasing the speed ratio can help generate steered swarms, excessively fast target-directed movements can lead to ``swarm-splitting'', i.e. leaders that pull free from followers and leave them stranded. This phenomenon is observed in Figure \ref{BiasesSetA}B4 at around $T\approx 290$, or in Figure \ref{BiasesSetC}B3 around $T\approx 80$. Under periodic boundary conditions, runaway leaders eventually reconnect with the stranded followers, leading to a periodic cycle (see the illustrative example of Figure \ref{BiasesSetC}B3). Of course, in a real-world scenario,  leaders would simply leave followers behind. 

Representative swarm behaviours under modulation of {\bf bias 3}, i.e. where we modulate the relative conspicuous of leaders moving towards or away from the destination, are shown in Figures \ref{BiasesSetA}C and Figure \ref{BiasesSetC}C. Notably, this form of bias was found to consistently generate a steered swarm in the target direction, over all tested ranges of  $\alpha_+/\alpha_-$ and for both two parameter regimes.

As a final exploration we examined swarm movement with all biases applied simultaneously: typical results are shown in Figure \ref{BiasesSetA}D for the high attraction/high alignment regime only. The application of multiple biases appears to reinforce steered movement in the direction of the destination, for example overturning the counter-directed swarms obtained for lower $\varepsilon$ and ratios of $\beta_+/\beta_-$. Yet escaping leaders can still occur when $\beta_+/\beta_-$ becomes too large, e.g. see Figure \ref{BiasesSetA}D4.

\section{Discussion}

Collective migration occurs when a population fashioned from interacting individuals self-organise and move in coordinated fashion. Recently, much attention has focused on the presence of leaders and followers, essentially a division of the group into distinct fractions that are either informed and aim to steer or naive and require steering, \cite{franks2002information,reebs2000can,mueller2013social}. Here we have formulated a continuous model to understand such phenomena, a non-local hyperbolic PDE system that explicitly incorporates separate leader and follower populations that have distinct responses to other swarm members. We considered distinct mechanisms through which leaders attempt to influence the swarm. Specifically, taking inspiration from the guidance provided by scout bees, \cite{Seeley,lindauer1957communication}, we focused on three different mechanisms: a bias in the leader alignment according to the target (\textbf{bias 1}), higher speed (\textbf{bias 2}) when moving towards the target and greater conspicuousness (\textbf{bias 3}) when moving towards the target.

We initially focused on simpler models of greater analytical tractability. First, a 100\% leader model composed only of informed members. Here only biases 1 or 2 operate, both proving effective at steering the group towards the destination. Maintaining group cohesion is, unsurprisingly, contingent on sufficiently strong attraction. Second, we considered a 100\% follower model: population members were naive but received some alignment bias, e.g. due to an implicitly present leader population. The range of dynamics generated by this model is more complicated, stemming from the more sophisticated alignment response. Nevertheless, introduction of bias acts to break the symmetry, significantly favouring the target direction. 

The full follower-leader model is capable of forming and maintaining a swarm that is consistently steered towards the destination, across a broad range of parameter regimes. Nevertheless, when examined individually, the distinct biases reveal varying levels of success at generating a steered swarm. First, biases 1 and 2 can, somewhat surprisingly, generate a swarm that moves away from the target, even inside plausible parameter regimes. Second, introduction of biases can lead to eventual dispersal of the swarm, pushing the system outside the regime in which attraction maintains swarm cohesion. Third, significant variation in speeds can lead to swarm splitting, where leaders split away from the group and leave followers stranded. Distinctly conspicuous leaders, however, consistently generated swarms moving towards the target, although we acknowledge the generality of this statement is limited by the purely numerical nature of the study. 

The varying success of the different influence strategies may stem from the manner in which the biases act. Biases 1 and 2 only {\em indirectly} influence followers: they describe behaviours in which the leaders alter their response according to the target direction, but do not directly enter the dynamics of the follower population. Any influence they exert on followers is therefore through altering the distribution of the leader population with respect to the followers, e.g. a variation in velocity that tends to polarise the position of leaders. 

Furthermore, the guidance efficiency provided by biases 1 and 2 appears to be related to the interactions parameter regime. On one hand, we observed counter-target directed swarms under bias 1 in the strong attraction-weak alignment regime. From a mathematical perspective this is reasonable, as the alignment strength ($q_l$) is directly proportional to the orientation bias ($\varepsilon$), see Eqs. \ref{Qlv_full_model} and \ref{Qal_M2}. We therefore speculate that bias 1 demands a sufficiently strong alignment. Otherwise, attraction dominates and the swarming group is directed accordingly, potentially against the target, as in Figure \ref{BiasesSetC}A1. On the other hand, bias 2 can lead to swarming against the target under strong alignment-strong attraction. In support of this, we remark that swarming direction derives from the transport terms (depending on $\beta_+:\beta_-$) and the competing social interactions. In summary, we speculate that bias 1 is favoured by interactions (specifically, alignment) while bias 2 is hindered by them. Bias 3, however, {\em directly} influence the followers: weighting the follower alignment to favour the target direction. For swarming populations it is worth stressing that these various biases may well act in concert: for example, in the context of bee swarms, speed variation may not be the intended mechanism for generating movement towards the target, it may rather be a side effect of altering the conspicuousness of nest-oriented scouts. 

As noted above, biases act to alter the relative positions of leader and follower populations, subtly weighting the interactions to break the symmetry of the system. Different parameter regimes lead to different follower/leader distributions, which we broadly classify as pull or push systems: in the former, leaders adopt a position at the front of the swarm, pulling the followers towards the target; in the latter, leaders are primarily concentrated in the rear, pushing them towards the destination. Transitions in the follower-leader distribution are highly contingent on parameter choice: for example, in Figure \ref{BiasesSetA}(C2-C3) we observed a sharp transition in the follower-leader distribution under a marginal variation in conspicuousness, in turn generating a significantly faster swarm. A more detailed analytical investigation into these transitions would be of significant interest, but lies outside the scope of the present study.

The model here provides substantial insight into the mechanisms through which informed leaders direct a swarm, yet its complexity has demanded certain simplifications. For example, in this preliminary work we have restricted to fixed leader and follower populations -- a reasonable approximation for, say, bee swarms, where a fixed subset of the population has explicit knowledge of the destination. In other instances, follower-leader distinction may be less clearcut and potentially transferable: for example, within cell populations ``leadership'' may be a chemically acquired characteristic determined by signals transmitted by other cells or the environment, \cite{atsuta2015fgf8}. Extensions of the model in this direction would require additional terms that account for the transfer between follower and leader status. We also note that the model here has focused on a simplified one-dimensional framework, though real-life collective migration phenomena are typically two or three dimensional in structure (for example, in bee swarms the streaker leaders adopt a position concentrated towards the upper portion of a 3D swarm). Further potential adaptations could include incorporating environmental heterogeneity, such as the need to overcome environmental obstacles, or modelling other forms of interaction, such as ``chase-and-run'' phenomena in which one population attempts to escape a population of pursuers. The latter is certainly relevant in ecological instances, for instance predator-prey relationships, but extends to various cellular populations including neural crest and placode cells \cite{theveneau2013chase}. While discrete models have been formulated to describe such processes, e.g. \cite{colombi2020}, a complementary continuous approach may yield further insights.
Finally, the model lends itself to studying  decision-making, i.e. where swarming may lead to consensus where there are multiple informed leader populations each exhibiting their own preferred direction. A fundamental contribution in this direction has been provided in \cite{couzin2005effective} through a discrete description.
Notwithstanding its simplifications, we believe the model presented here provides a starting point for future investigations into the role of heterogeneity on collective migration phenomena.

\subsection*{Acknowledgements}
This research was partially supported by the Italian Ministry of Education, University and Research (MIUR) through the ``Dipartimenti
di Eccellenza'' Programme (2018-2022) -- Dipartimento di Scienze Matematiche ``G. L. Lagrange'', Politecnico di Torino
(CUP: E11G18000350001). SB is member of GNFM (Gruppo Nazionale per la Fisica Matematica) of INdAM (Istituto
Nazionale di Alta Matematica), Italy.

\appendix

\section{Fixed Parameters} \label{fixedparameters}

The parameters in Table \ref{table2} are common to all models and set at a fixed reference value, in accordance with those in \cite{Raluca}.

\begin{table}[h!] 
\begin{center}
\caption{Table of parameters with fixed values.}
\label{table2}       
\begin{tabular}{llll}
\hline\noalign{\smallskip}
Parameter     &Description        &Value [Unit]    \\
\noalign{\smallskip}\hline\noalign{\smallskip}
$y_0$         &shift of the turning function    &$2$      \\   
$s_r$         &attraction range    &0.25 [L]    \\   
$s_{l}$         &alignment range    &0.5 [L]     \\  
$s_a$         &attraction range    &1 [L]      \\   
$m_r$         &width of repulsion kernel    &0.25/8 [L]      \\  
$m_{l}$         &width of alignment kernel    &0.5/8 [L]      \\ 
$m_a$         &width of attraction kernel    &1/8 [L]      \\ 
$L$         &domain length    &10 [L]\\
$\gamma$            &follower speed &0.1 [L/T]  \\ 
\noalign{\smallskip}\hline
\end{tabular}
\end{center}
\end{table}

\section{Numerical Method} \label{numericalmethods}

Numerical simulations are performed on a 1D spatial domain $[0,L]$ with periodic boundary conditions imposed at $x = 0, L$: 
\[
u^+(0,t) = u^+(L,t)\,, \quad u^-(0,t)  = u^-(L,t)\,, 
\] 
and similarly for $v^{\pm}$. The numerical scheme invokes a Methods of Lines approach, where initial discretisation over space yields a system of ordinary differential equations that are subsequently integrated over time. Spatial movement terms are approximated with a first order upwind scheme. Switching terms demand approximation of the infinite attraction/repulsion/ alignment integrals. For this we exploit the Gaussian nature of the Kernel functions and approximate the integrals on finite domains $[0,6i]$, $i=s_r, s_a$, for attractive and repulsive kernels and $[0,2s_{l}]$ for alignment. The integral itself is approximated using Simpson's method. We finally remark that under the periodic boundary conditions the integrals are wrapped around the domain.

\section{Stability analysis for follower-only model}

\subsection{Expressions for stability analysis} \label{follower_only_lin_stab_appendix}

Basic algebra provides the following expressions for $\lambda^{i^\pm}_{i^\pm}$, for $i \in\left\{u,v\right\}$:
\begin{eqnarray}
\label{derivate_lambda}
\lambda^{u^+}_{u^-}&= 0.5 \lambda_2  \left[1-\tanh^2(q_l (u^{**}-u^*)-y_0) \right]
\left( q_r \hat{K}_r^+(k)-q_a \hat{K}_a^+(k) + q_l \hat{K}_{l}^+(k) \right)\,, \\
\lambda^{u^-}_{u^-}  &=  0.5 \lambda_2  \left[1-\tanh^2(-q_l (u^{**}-u^*)-y_0) \right]
\left( q_r \hat{K}_r^-(k)-q_a \hat{K}_a^-(k) - q_l \hat{K}_{l}^+(k) \right)\,, \nonumber\\
\lambda^{u^+}_{u^+} &= 0.5 \lambda_2  \left[1-\tanh^2(q_l (u^{**}-u^*)-y_0) \right]
\left( q_r \hat{K}_r^+(k)-q_a \hat{K}_a^+(k) - q_l \hat{K}_{l}^-(k) \right)\,, \nonumber\\  
\lambda^{u^-}_{u^+} &=0.5 \lambda_2  \left[1-\tanh^2(-q_l (u^{**}-u^*)-y_0) \right]
\left( q_r \hat{K}_r^-(k)-q_a \hat{K}_a^-(k) + q_l \hat{K}_{l}^-(k) \right)\,, \nonumber
\end{eqnarray}
where
$\hat{K}_{j}^\pm (k), j=r,a,l$ denote the Fourier transform of the kernel $K_j(s)$, i.e.
\[
\hat{K}_{j}^\pm (k)= \int_{-\infty}^{+\infty} K_j(s) e^{\pm iks} ds=\exp \left( \pm i s_j k-\frac{k^2 m_{l}^2}{2} \right), \quad j=r,a,l.
\]
Substituting Eq. \ref{derivate_lambda} into Eqs. \ref{C} and \ref{D} yields
\begin{eqnarray*}
C(k) =&-2 \lambda_1 -\lambda_2 - 0.5 \lambda_2 \left[ \tanh(q_l(u^{**}-u^*- \eta)-y_0) + \tanh(-q_l(u^{**}-u^*- \eta)-y_0) \right] \\
&+q_l (\hat{K}_{l}^+(k) + \hat{K}_{l}^-(k)) \{ u^* \left[ 0.5 \lambda_2 (1-\tanh^2 (q_l(u^{**}-u^*- \eta)-y_0)) \right]  \\
&+u^{**} \left[ 0.5 \lambda_2 (1-\tanh^2 (-q_l(u^{**}-u^*- \eta)-y_0)) \right] \} \\
D(k) = & 4 \gamma^2 k^2 + 2 \gamma \lambda_2 ik  \{   \tanh (-q_l(u^{**}-u^*- \eta)-y_0)
-\tanh (q_l(u^{**}-u^*- \eta)-y_0)  \\ 
&+u^* [ 1-\tanh^2 (q_l(u^{**}-u^*- \eta)-y_0) ](-2 q_r \hat{K}_{r}^+(k) +2q_a \hat{K}_{a}^+(k) -q_l \hat{K}_{l}^+(k) +q_l \hat{K}_{l}^-(k))\\
&+ u^{**} [ 1-\tanh^2 (-q_l(u^{**}-u^*- \eta)-y_0) ](2 q_r \hat{K}_{r}^-(k) -2q_a \hat{K}_{a}^-(k) -q_l \hat{K}_{l}^+(k) +q_l \hat{K}_{l}^-(k)) \}.
\end{eqnarray*}

\subsection{Steady state variation in parameter space}\label{follower_only_steadystates}

\begin{figure}
    \centering
    {\includegraphics[width=\textwidth]{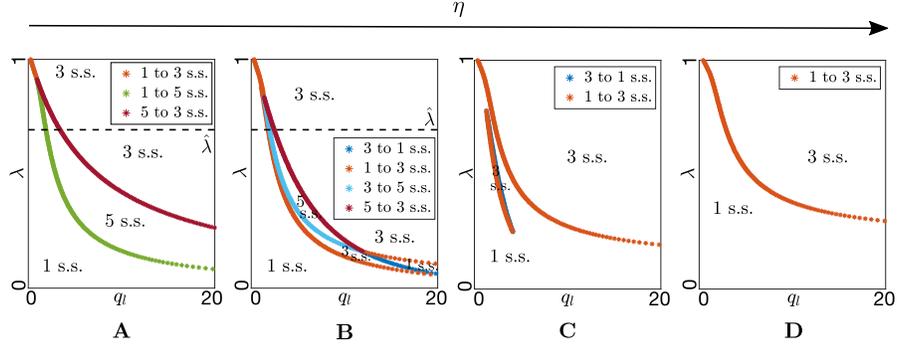}}
    \caption{Two parameter bifurcation diagram in ($q_l,\lambda$) space, obtained for (A) $\eta=0$, (B) $\eta=0.04$, (C) $\eta=0.2$, (D) $\eta=0.4$. Dashed lines in (A,B) indicate the value of $\lambda$, denoted as $\hat{\lambda}$, resulting from $\lambda_1=0.2$ and $\lambda_2=0.9$ (compare to bifurcation plots in Fig. \ref{FO_model1}A and B). The mean total follower density is set as $A_u=2$.}
    \label{135_ss_qal}
\end{figure}

When alignment impacts on the social interactions, i.e. $q_l \neq 0$, Eq. \ref{h1=0} can have one, three or five solutions, depending on the value of $\lambda$ and $\eta$.
Specifically, for smaller $\eta$ two-parameter numerical bifurcation diagrams in $(q_l,\lambda)$ space indicate a threshold value $\lambda^{*} \in (0,1)$ such that if $\lambda > \lambda^*$ then there are up to three solutions. Conversely, if $\lambda < \lambda^*$ there are up to five solutions (Figure \ref{135_ss_qal}A,\ref{135_ss_qal}B). 
As $\eta$ increases, the parameter region resulting in $5$ steady states reduces, completely disappearing for $\eta = 0.2$, see Figures \ref{135_ss_qal}A, \ref{135_ss_qal}B, \ref{135_ss_qal}C. For $\eta=0.4$, Eq. \ref{h1=0} shows one or three solutions, see Figure \ref{135_ss_qal}D.


\subsection{Effect of alignment on equilibrium points}\label{alignment_equilibrium_points}

\begin{figure}
    \centering
    {\includegraphics[width=\textwidth]{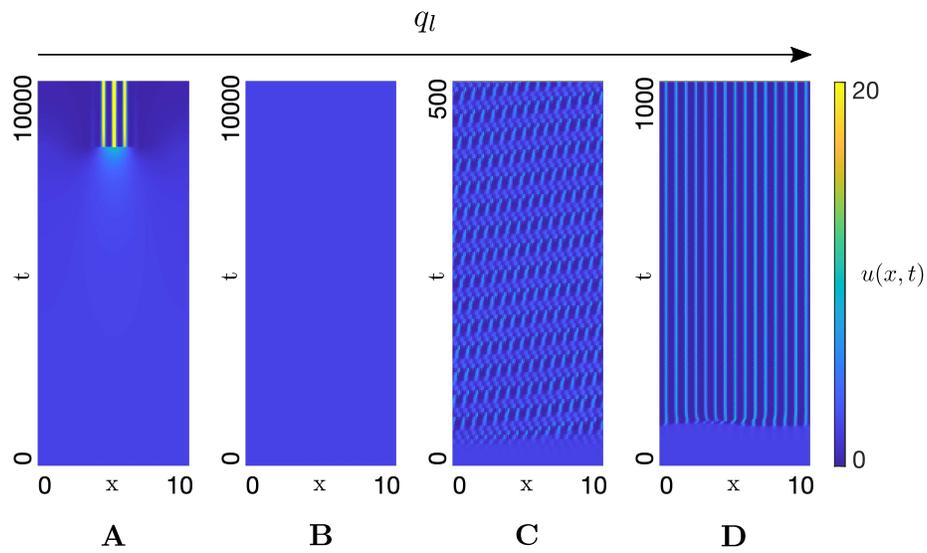}}
    \caption{Pattern formation for the follower-only model under the unbiased scenario, i.e. $\eta=0$, obtained for (A) $q_l=0.4$, (B) $q_l=1.3$, (C) $q_l=2$, (D) $q_l=2.5$. The other parameters are set as $q_r=0, q_a=0.25, A_u=2, y_0=2,  \lambda_1=0.8, \lambda_2=3.6$. ICs are perturbations of $(u^*, u^{**})=(1,1)$ in case (A), (B) and (D). In case (C), ICs are perturbations of $(u^*, u^{**})=(1.62,0.38)$.}
    \label{simu_bif_qal_follower_only}
\end{figure}

The bifurcation diagram in Figure \ref{FO_model1}A shows that in the absence of any external bias,various stationary and temporal patterns can emerge, as previously described \cite{Raluca}. For example, under weak alignment and sufficiently strong attraction, pattern formation occurs whereby the population aggregates through mutual attraction (Figure \ref{simu_bif_qal_follower_only}A). Under weak alignment, however, a symmetry is maintained between the proportions of ($\pm$) populations and the aggregate remains stationary.
For stronger alignment the HSS becomes stable and the population remains uniformly dispersed across the domain, see Figure \ref{simu_bif_qal_follower_only}B, yet
for even stronger alignment two symmetric branches appear with both locally undergoing a saddle point bifurcation and leading to the formation of moving patterns (Figure \ref{simu_bif_qal_follower_only}C). Note, though, for parameters in this region and quasi-symmetrically distributed initial populations, patterns are equally likely to favour the $(+)$ or $(-$) directions. Within this alignment range, the central equilibrium turns back to generate stationary aggregations (Figure \ref{simu_bif_qal_follower_only}D) and when the system crosses a critical value a pitchfork bifurcation arises: the solution $(u^*, u^{**})=(1,1)$ loses its stability in the homogeneous space and the population jumps to one of the stable branches.

\clearpage
\newpage


\bibliographystyle{plain}
\bibliography{Reference}   

\begin{thebibliography}{10}

\bibitem{armstrong2006}
N.~J. Armstrong, K.~J. Painter, and J.~A. Sherratt.
\newblock A continuum approach to modelling cell--cell adhesion.
\newblock {\em Journal of Theoretical Biology}, 243(1):98--113, 2006.

\bibitem{atsuta2015fgf8}
Y.~Atsuta and Y.~Takahashi.
\newblock {FGF}8 coordinates tissue elongation and cell epithelialization
  during early kidney tubulogenesis.
\newblock {\em Development}, 142(13):2329--2337, 2015.

\bibitem{beekman2006does}
M.~Beekman, R.~L. Fathke, and T.~D. Seeley.
\newblock How does an informed minority of scouts guide a honeybee swarm as it
  flies to its new home?
\newblock {\em Animal Behaviour}, 71(1):161--171, 2006.

\bibitem{bernardi2018particle}
S.~Bernardi, A.~Colombi, and M.~Scianna.
\newblock A particle model analysing the behavioural rules underlying the
  collective flight of a bee swarm towards the new nest.
\newblock {\em Journal of Biological Dynamics}, 12(1):632--662, 2018.

\bibitem{brent2015ecological}
L.~J.~N. Brent, D.~W. Franks, E.~A. Foster, K.~C. Balcomb, M.~A. Cant, and
  D.~P. Croft.
\newblock Ecological knowledge, leadership, and the evolution of menopause in
  killer whales.
\newblock {\em Current Biology}, 25(6):746--750, 2015.

\bibitem{Carrillo}
J.~A. Carrillo, M.~Fornasier, G.~Toscani, and F.~Vecil.
\newblock Particle, kinetic, and hydrodynamic models of swarming.
\newblock In {\em Mathematical modeling of collective behavior in
  socio-economic and life sciences}, pages 297--336. Springer, 2010.

\bibitem{chen2020}
L.~Chen, K.~J. Painter, C.~Surulescu, and A.~Zhigun.
\newblock Mathematical models for cell migration: a non-local perspective.
\newblock {\em Philosophical Transactions of the Royal Society B},
  375(1807):20190379, 2020.

\bibitem{cheung2013collective}
K.~J. Cheung, E.~Gabrielson, Z.~Werb, and A.~J. Ewald.
\newblock Collective invasion in breast cancer requires a conserved basal
  epithelial program.
\newblock {\em Cell}, 155(7):1639--1651, 2013.

\bibitem{colombi2017discrete}
A.~Colombi, M.~Scianna, and A.~Alaia.
\newblock A discrete mathematical model for the dynamics of a crowd of gazing
  pedestrians with and without an evolving environmental awareness.
\newblock {\em Computational and Applied Mathematics}, 36(2):1113--1141, 2017.

\bibitem{colombi2020}
A.~Colombi, M.~Scianna, K.~J. Painter, and L.~Preziosi.
\newblock Modelling chase-and-run migration in heterogeneous populations.
\newblock {\em Journal of Mathematical Biology}, 80(1):423--456, 2020.

\bibitem{couzin2005effective}
I.~D. Couzin, J.~Krause, N.~R. Franks, and S.~A. Levin.
\newblock Effective leadership and decision-making in animal groups on the
  move.
\newblock {\em Nature}, 433(7025):513--516, 2005.

\bibitem{couzin2002collective}
I.~D. Couzin, J.~Krause, R.~James, G.~D. Ruxton, and N.~R. Franks.
\newblock Collective memory and spatial sorting in animal groups.
\newblock {\em Journal of Theoretical Biology}, 218(1):1--12, 2002.

\bibitem{dingle2014migration}
H.~Dingle.
\newblock {\em Migration: the biology of life on the move}.
\newblock Oxford University Press, USA, 2014.

\bibitem{diwold2011deciding}
K.~Diwold, T.~M. Schaerf, M.~R. Myerscough, M.~Middendorf, and M.~Beekman.
\newblock Deciding on the wing: in-flight decision making and search space
  sampling in the red dwarf honeybee apis florea.
\newblock {\em Swarm Intelligence}, 5(2):121--141, 2011.

\bibitem{eftimie2012}
R.~Eftimie.
\newblock Hyperbolic and kinetic models for self-organized biological
  aggregations and movement: a brief review.
\newblock {\em Journal of Mathematical Biology}, 65(1):35--75, 2012.

\bibitem{Raluca}
R.~Eftimie, G.~De~Vries, M.~A. Lewis, and F.~Lutscher.
\newblock Modeling group formation and activity patterns in self-organizing
  collectives of individuals.
\newblock {\em Bulletin of Mathematical Biology}, 69(5):1537, 2007.

\bibitem{fetecau2012mathematical}
R.~C. Fetecau and A.~Guo.
\newblock A mathematical model for flight guidance in honeybee swarms.
\newblock {\em Bulletin of Mathematical Biology}, 74(11):2600--2621, 2012.

\bibitem{franks2002information}
N.~R. Franks, S.~C. Pratt, E.~B. Mallon, N.~F. Britton, and D.~J.~T. Sumpter.
\newblock Information flow, opinion polling and collective intelligence in
  house--hunting social insects.
\newblock {\em Philosophical Transactions of the Royal Society B},
  357(1427):1567--1583, 2002.

\bibitem{friedl2009collective}
P.~Friedl and D.~Gilmour.
\newblock Collective cell migration in morphogenesis, regeneration and cancer.
\newblock {\em Nature Reviews Molecular Cell Biology}, 10(7):445--457, 2009.

\bibitem{friedl2012classifying}
P.~Friedl, J.~Locker, E.~Sahai, and J.~E. Segall.
\newblock Classifying collective cancer cell invasion.
\newblock {\em Nature Cell Biology}, 14(8):777--783, 2012.

\bibitem{greggers2013scouts}
U.~Greggers, C.~Schoening, J.~Degen, and R.~Menzel.
\newblock Scouts behave as streakers in honeybee swarms.
\newblock {\em Naturwissenschaften}, 100(8):805--809, 2013.

\bibitem{helbing1995social}
D.~Helbing and P.~Molnar.
\newblock Social force model for pedestrian dynamics.
\newblock {\em Physical Review E}, 51(5):4282, 1995.

\bibitem{ioannou2015potential}
C.~C. Ioannou, M.~Singh, and I.~D. Couzin.
\newblock Potential leaders trade off goal-oriented and socially oriented
  behavior in mobile animal groups.
\newblock {\em The American Naturalist}, 186(2):284--293, 2015.

\bibitem{janson2005honeybee}
S.~Janson, M.~Middendorf, and M.~Beekman.
\newblock Honeybee swarms: how do scouts guide a swarm of uninformed bees?
\newblock {\em Animal Behaviour}, 70(2):349--358, 2005.

\bibitem{kretz2006experimental}
T.~Kretz, A.~Gr{\"u}nebohm, M.~Kaufman, F.~Mazur, and M.~Schreckenberg.
\newblock Experimental study of pedestrian counterflow in a corridor.
\newblock {\em Journal of Statistical Mechanics: Theory and Experiment},
  2006(10):P10001, 2006.

\bibitem{lindauer1957communication}
M.~Lindauer.
\newblock Communication in swarm-bees searching for a new home.
\newblock {\em Nature}, 179(4550):63--66, 1957.

\bibitem{mogilner1999}
A.~Mogilner and L.~Edelstein-Keshet.
\newblock A non-local model for a swarm.
\newblock {\em Journal of Mathematical Biology}, 38(6):534--570, 1999.

\bibitem{mueller2013social}
T.~Mueller, R.~B. O'Hara, S.~J. Converse, R.~P. Urbanek, and W.~F. Fagan.
\newblock Social learning of migratory performance.
\newblock {\em Science}, 341(6149):999--1002, 2013.

\bibitem{reebs2000can}
S.~G. Reebs.
\newblock Can a minority of informed leaders determine the foraging movements
  of a fish shoal?
\newblock {\em Animal Behaviour}, 59(2):403--409, 2000.

\bibitem{schultz2008}
K.~M. Schultz, K.~M. Passino, and T.~D. Seeley.
\newblock The mechanism of flight guidance in honeybee swarms: subtle guides or
  streaker bees?
\newblock {\em Journal of Experimental Biology}, 211(20):3287--3295, 2008.

\bibitem{schumacher2016}
L.~J. Schumacher, P.~M. Kulesa, R.~McLennan, R.~E. Baker, and P.~K. Maini.
\newblock Multidisciplinary approaches to understanding collective cell
  migration in developmental biology.
\newblock {\em Open Biology}, 6(6):160056.

\bibitem{Seeley}
T.~D. Seeley.
\newblock {\em Honeybee democracy}.
\newblock Princeton University Press, 2010.

\bibitem{Sumpter}
D.~J.~T. Sumpter.
\newblock {\em Collective animal behavior}.
\newblock Princeton University Press, 2010.

\bibitem{theveneau2013chase}
E.~Theveneau, B.~Steventon, E.~Scarpa, S.~Garcia, X.~Trepat, A.~Streit, and
  R.~Mayor.
\newblock Chase-and-run between adjacent cell populations promotes directional
  collective migration.
\newblock {\em Nature Cell Biology}, 15(7):763--772, 2013.

\bibitem{topaz2006}
C.~M. Topaz, A.~L. Bertozzi, and M.~A. Lewis.
\newblock A nonlocal continuum model for biological aggregation.
\newblock {\em Bulletin of Mathematical Biology}, 68(7):1601, 2006.

\bibitem{turing1952}
A.~Turing.
\newblock The chemical basis of morphogenesis.
\newblock {\em Philosophical Transactions of the Royal Society B}, 237:37--72,
  1952.

\bibitem{vishwakarma2018mechanical}
M.~Vishwakarma, J.~Di~Russo, D.~Probst, U.~S Schwarz, T.~Das, and J.~P. Spatz.
\newblock Mechanical interactions among followers determine the emergence of
  leaders in migrating epithelial cell collectives.
\newblock {\em Nature Communications}, 9(1):1--12, 2018.

\bibitem{westley2018collective}
P.~A.~H. Westley, A.~M. Berdahl, C.~J. Torney, and D.~Biro.
\newblock Collective movement in ecology: from emerging technologies to
  conservation and management.
\newblock {\em Philosophical Transactions of the Royal Society B},
  237:20170004, 2018.

\end{thebibliography}

%
%

\end{document}